\documentclass[aps,pra,floatfix,twocolumn,nofootinbib]{revtex4-2}
\usepackage{bm}
\usepackage[dvips]{epsfig}
\usepackage{graphicx} 
\usepackage{amsmath} 
\usepackage{amssymb}

\usepackage{xcolor}
\usepackage[normalem]{ulem}
\def\bea{\begin{eqnarray}}
\def\eea{\end{eqnarray}}

\newcommand{\comment}[1]{}
\newcommand{\BEQ}{\begin{align}}
\newcommand{\EEQ}{\end{align}}
\newcommand{\BEA}{\begin{eqnarray}}
\newcommand{\EEA}{\end{eqnarray}}
\newcommand{\nn}{\nonumber \\}

\renewcommand{\d}{{\rm d}}

\newcommand{\ep}{\varepsilon}

\newcommand{\by}{\boldsymbol{y}}

\newcommand{\bk}{{\bf k}}

\newcommand{\bq}{\boldsymbol{q}}

\newcommand{\pa}{\partial}
\newcommand{\om}{\omega}

\begin{document}

\title{Enhanced detection of time-dependent dielectric structure:\\
Rayleigh limit and quantum vacuum }

\author{ V. E. Mkrtchian, H. S. Avetisyan, and A. E. Allahverdyan}
\affiliation{
Alikahanyan National Laboratory (Yerevan Physics Institute), 2 Alikhanyan
Brothers Street, Yerevan 0036, Armenia}

\begin{abstract}
Detection of scattered light can determine the susceptibility of dielectrics. Such imaging normally holds Rayleigh's limit: details finer than the wavelength of the incident light cannot be determined from the far-field zone. We show that time-modulation of an inhomogeneous dielectric can be used to determine its susceptibility. To this end, we focus on the inverse quantum optics problem for spatially and temporally modulated metamaterial, whose dielectric susceptibility is similar to moving dielectrics. We show that the vacuum contribution to the photodetection signal is non-zero due to the negative frequencies even in the far-field zone. {Hence, certain dielectric features can be determined without radiating any incident field on the dielectric. When the incident light is scattered (or re-radiated), the determination of dielectric susceptibility is enhanced and overcomes  Rayleigh's limit in the far-field zone. We study similar effects for an inhomogeneous dielectric moving with a constant speed, a problem we consider within relativistic optics. Now the vacuum contribution to the photodetection signal reflects dielectric features, can be long-range in space but is not a far-field effect. }

\end{abstract}

\comment{Letter to PRL editors (100 words).

We made a new proposal for improving Rayleigh resolution limit of inverse optics, where the dielectric susceptibility is determined by shining light on a sample and detecting the scattered light. Our proposal applies to samples with a time-dependent susceptibility. Two mechanisms we explored were moving samples and space-time-modulated metamaterials. We show that quantum vacuum effects in photodetection can determine certain features of susceptibility even without shining any light. When the shining is there, our mechanism overcomes classical Rayleigh limit. Our results could advance non-invasive material characterization techniques. They are timely given the current interest in metrology, metamaterials and Doppler metrology.

}

\maketitle

\section{Introduction}

Inverse optics determines the dielectric susceptibility of inhomogeneous materials by irradiating them with (incident) light with known characteristics and detecting scattered light \cite{baltes,Devaney}. Recent advances in inverse optics relate to the use of quantum features of incident light \cite{Shih,Abouraddy,Dowling,monken,liu,we}. In particular, quantum features improve on the classical Rayleigh limit that bounds the far-field resolution of dielectric susceptibility, given the wavelength of the incident light. Improvement is possible (up to eight times for two-photon light \cite{we}), but it connects with difficulties of preparing specific quantum states of light and
with having prior information on the dielectric sample \cite{we}. 

We are looking for additional resources that will allow for better resolution. Our results identify a resource that relies on space-time modulation of an inhomogeneous dielectric. As a result of the modulation, the dielectric has a space-time-dependent susceptibility similar to that of a moving dielectric. Our proposal is dual to Doppler metrology \cite{doppler_metro}: we set the moving object in motion to analyze its internal structure, rather than to find the velocity of the object.

The problem of moving dielectric has a fundamental appeal because it gave rise to special relativity \cite{albert,hermann,pal}. Progress in this field was steady and impressive: the Fresnel-Fizeau drag, the Doppler effect(s), the relativistic Snell--Descartes law, Cherenkov radiation \cite{tamm}, light amplification via moving mirrors \cite{kiefer}, {\it etc}. Quantization of the electromagnetic field in the presence of moving dielectric media \cite{Matloob1, Matloob2,Horsley} led to more recent results; e.g. the quantum friction phenomenon \cite{Horsley,vem, pendry1}. 

\comment{it was shown that the Hamiltonian of the coupled matter-field system contains negative-energy normal modes which are the origin of the quantum friction phenomenon}

Electrodynamics of relativistically moving bodies has traditionally focused on electron clouds or plasma jets \cite{boloto, bulanov, kiefer}. A recent activity in creating metamaterials with a space-time-modulated electric susceptibility renewed interest in this field \cite{galiffi,pendry,caloz}. Though these metamaterials do not move, they to a certain extent mimic moving dielectrics.

Here we aim to show that motion or time modulation of an inhomogeneous dielectric is a resource for improving
the resolution of its internal structure, i.e. seeing deeper into the dielectric structure. It turns out that for a modulated or moving dielectric, incident light is not even necessary: the object reveals its structure through the vacuum response of the photodetector. This effect comes from up-conversion of field modes/operators from negative to positive frequencies. Specifically, we show that the vacuum response of the far-field photodetector can be used to determine the characteristic length of the dielectric susceptibility for moving dielectrics. 

{In section \ref{modulated}, we develop scattering theory for a spatially and temporally modulated metamaterial, whose dielectric susceptibility is similar to that of moving dielectrics. We show that in this case there is a vacuum response (the one taking place without any incident light), and that this response can take place in the far-field zone of spherical waves. In its specific realizations (broadband harmonic modulation) it has certain analogies to the anomalous Doppler effect; see \cite{ginzburg,ginzburg1996} for reviews on this effect. When incident light and its scattering are present, we show that far-field detection can overcome the classical Rayleigh resolution limit. This allows for an enhanced detection of dielectric features along the direction of modulation. Technically, this means that the Fourier image of the dielectric susceptibility is determined via the far-field photodetection (which measures the intensity of the quantum electric field) at wave vectors sizably larger in modulus than $\om/c$, where $\om$ is the photodetection frequency. }

The above results concern the space-time modulated dielectric, a behavior that mimics rectilinear motion. To understand precise relations between these two situations, in section \ref{moving} we study -- within relativistic optics -- a dielectric moving with a constant velocity. Here the vacuum photodetection is limited to a near-field zone of cylindrical waves, but it can be a long-range effect if the motion is ultra-relativistic. {Long-range means that the signal intensity decays as a power law (not exponent) with the distance from the trajectory of the moving dielectric. When cylindrical waves are in the far-field, their intensity decays as the inverse of the distance.}

This paper is organized as follows. Sections \ref{maxwell} and \ref{modulated} study quantum scattering theory for a space-time modulated dielectric. The vacuum response is studied in section \ref{vaco}. In section \ref{shino} we research the modulated situation under one-photon irradiation. Section \ref{moving} develops a relativistic scattering theory for a moving dielectric, accounting for both spatial and temporal dispersion (though we eventually focus on temporal dispersion only). In contrast to the phenomenological approach of sections \ref{maxwell} and \ref{modulated}, the relativistic consideration of section \ref{moving} is based on first principles. Section \ref{moving} can be read independently from sections \ref{maxwell} and \ref{modulated}. The last section summarizes.

\comment{
A recent activity in creating metamaterials with space-time modulated electric
and/or magnetic permeabilities has iniciated interest to manifestation of its
pecularities in traditional experiments of light scattering.

\textbf{Superscattering }PRL \textbf{122}, 063901 (2019)

A four-fold increase in the scattering upper limit has been
observed$\rightarrow$highly-efficient antennas for harvesting energy.
}

\section{ Scattering theory for inhomogeneous dielectrics}
\label{maxwell}

We develop quantum light scattering theory in space-time modulated isotropic, inhomogeneous dielectric
metamaterials embedded in vacuum. We work within quantum macroscopic electrodynamics, 
where the inhomogeneous, time-dependent dielectric is described via 
susceptibility $\varepsilon\left(  \mathbf{r,}t\right)-1$ (no magnetic features), and the standard quantities for the electromagnetic field \cite{landau}: 
\BEA
\mathbf{D}\left(  \mathbf{r,}t\right)  =\varepsilon\left(  \mathbf{r,}
t\right)  \mathbf{E}\left(  \mathbf{r,}t\right), ~~\mathbf{H}\left(
\mathbf{r,}t\right) = \mathbf{B}\left(  \mathbf{r,}t\right).
\label{tk}
\EEA
We use Gaussian units with $c=\hbar=\epsilon_0=1$.
Employing (\ref{tk}) in Maxwell equations and excluding $\mathbf{B}$ leads to the Helmholtz equation for $\mathbf{E}$ \cite{landau}:
\begin{align}
\label{a1}
&\partial_\beta\partial_\beta E_{\alpha}-\partial_\alpha\partial_\beta E_{\beta}-\partial_{t}^{2}E_{\alpha}=\partial_{t}^{2}(\bar{\varepsilon} E_{\alpha}),\\
&\bar{\varepsilon}\left(  \mathbf{r,}t\right)  \equiv\varepsilon\left(
\mathbf{r,}t\right)  -1,\\ 
&\alpha,\beta=x,y,z, ~~~ \mathbf{r}=(x,y,z), 
\label{a2}
\end{align}
where repeated indices imply summation.

\comment{
\begin{eqnarray}
\label{a1}
\partial_{\alpha}\partial_{\beta}E_{\beta}+\partial_{t}^{2}\left[
\varepsilon\left(  \mathbf{r,}t\right)  E_{\alpha}\right]  -\partial_\beta\partial_\beta E_{\alpha}
=0,~~\alpha,\beta=1,2,3,
\end{eqnarray}
where $\pa_\alpha=\pa/\pa x_\alpha$, repeated indices imply summation, and $c=\hbar=\epsilon_0=1$.
The first Maxwell equation $\partial_{\alpha}\left[  \varepsilon\left(  \mathbf{r,}t\right)  E_{\alpha
}\right]  =0$, which is consistent with (\ref{a1}), leads to the separation of the 
"interaction potential" $\bar{\varepsilon}\left(  \mathbf{r,}t\right)  \equiv\varepsilon\left(
\mathbf{r,}t\right)  -1$ in (\ref{a1}):
}

We assume that the electromagnetic field is quantized making (\ref{a1}) the Heisenberg equation for the operator $E_\beta$ \cite{landau}. But the matter is classical and has the average dielectric susceptibility $\varepsilon\left(  \mathbf{r,}t\right)-1$ \cite{landau}. Hence, in (\ref{tk}) we left aside both the time-dispersion of dielectrics 
and its quantum features. Both are straightforward to include: time-dispersion will amount to an integral equation in (\ref{tk}), while the quantum features can be introduced by promoting $\varepsilon\left(  \mathbf{r,}t\right)$ tos an operator and postponing the averaging over the dielectric state till the final results \cite{landau}. We avoid introducing these complications because in the present phenomenological set-up, none of them are essential to our main results. For the same reason, we neglect anisotropy in $\varepsilon\left(  \mathbf{r,}t\right)$.
In section \ref{moving}, we study a relativistically moving dielectric, and we need to account (at least) for time-dispersion features. 

Eqs.~(\ref{tk}--\ref{a2}) apply to quantum optics, where intensities are measured via photodetection \cite{garrison}. A photodetector is localized around position $\mathbf{r}$, works at an atomic transition frequency $\omega>0 $ and measures e.g. the mean electric field intensity of the scattered radiation in the long-time limit \cite{garrison}
\begin{align}
I [\omega,\mathbf{r};|\psi\rangle\langle \psi|]=
\langle \psi| {E}_{\alpha}^\dagger[
\omega,\mathbf{r}]\,  {E}_{\alpha} [
\omega,\mathbf{r}]\,|\psi\rangle,~~\om>0, \label{11}%
\end{align}
where $|\psi\rangle$ is the (initial) quantum state of the field, $\dagger$ means hermitian conjugation, and
\begin{align}
\label{fou}
{E}_{\alpha}[\omega,\mathbf{r}] = \int\frac{\d t}{2\pi}e^{it\omega} {E}_{\alpha}(t,\mathbf{r})
\end{align}
is the Fourier transform of the electric field operator.

Within the standard setup of scattering $\mathbf{E}$ reads \cite{Devaney}:
\begin{align}
\mathbf{E=E}^{[\text{in}]  }+\mathbf{E}^{[\text{s}] },
\label{6}
\end{align}
where the incident field $\mathbf{E}^{[\text{in}] }(\mathbf{r},t)$ satisfies 
free Helmholtz's equation, i.e. nullifies the left-hand-side of (\ref{a1}). Hence $\mathbf{E}^{[\text{in}] }$ has the standard quantized representation \cite{garrison}:
\begin{align}
&\mathbf{{E}}^{[\text{in}]  }\left(  \mathbf{r,}t\right)  =\frac
{-1}{2\pi}\int \d\mathbf{q}\,\sqrt{q}\,\mathbf{e}_{\lambda}\left(  \mathbf{q}\right)
e^{i\left(  \mathbf{q\cdot r}-qt\right)  }{a}_{\lambda}\left(  \mathbf{q}%
\right)  +\text{h.c.},\label{Ert} \\
\label{808}%
& \mathbf{q\cdot e_\lambda(q)}=0,\quad \mathbf{e_\lambda(q) \cdot e_{\lambda'}(q)}=\delta_{\lambda\lambda'}, \quad \lambda=1,2, \\
\label{gabal}
& {\rm e}_{\lambda\alpha}\left(  \mathbf{q}\right)  {\rm e}_{\lambda\beta}\left(
\mathbf{q}\right)  =\delta_{\alpha\beta}-{q}_{\alpha}{q}_{\beta
}/q^2, \quad q=|\mathbf{q}|,  \\ \label{5}%
& 
{a}_{\lambda}\left(  \mathbf{q}\right)  
{a}_{\lambda^{\prime}}^{\dagger}\left(  \mathbf{q}^{\prime}\right)
-{a}_{\lambda^{\prime}}^{\dagger}\left(  \mathbf{q}^{\prime}\right)
{a}_{\lambda}\left(  \mathbf{q}\right)  
= \delta_{\lambda\lambda^{\prime
}}\delta\left(  \mathbf{q-q}^{\prime}\right), 
\end{align}
where $\mathbf{e}_{\lambda}\left(  \mathbf{q}\right)=\mathbf{e}_{\lambda}\left(  -\mathbf{q}\right)$ is the real
polarisation vector, $A+{\rm h.c.}=A+A^\dagger$, $\delta_{\lambda\lambda'}$ is the Kroenecker delta, and ${a}_{\lambda}(\mathbf{q})$ is the annihilation operator. Note that ${a}_{\lambda}(\mathbf{q})$ and ${a}_{\lambda'}(\mathbf{q}')$ commute. Eqs.~(\ref{808}, \ref{gabal}) define orthogonality features of polarization vectors, where repeated $\lambda$ implies summation. Now $\mathbf{{E}}^{[\text{in}]  }(\mathbf{r},t)$ contains contributions both from creation ${a}^\dagger_{\lambda}\left(  \mathbf{q}\right)$ and annihilation ${a}_{\lambda}\left(  \mathbf{q}\right)$ operators at (resp.) positive and negative frequencies, while only ${a}_{\lambda}\left(  \mathbf{q}\right)$ contribute to $\mathbf{{E}}^{[\text{in}]  }[\om>0,\mathbf{r}]$. 

To understand our results, it suffices to work within the first-order Born approximation, where the scattered field $\mathbf{E}^{[\text{s}] }[\om,\mathbf{r}]$ in (\ref{6}) is determined via taking the Fourier transform (\ref{fou}) of (\ref{a1}), and assuming that $\bar{\varepsilon}{E}_\alpha^{[\text{s}] }$ is small: 
\begin{align}
\label{a11}
 \{\partial_\beta\partial_\beta -& \partial_\alpha  \partial_\beta +\omega^{2}\}E_{\alpha}^{[\text{s}] }[\om,\mathbf{r}]\\
 &=-\omega^{2}\int\d\omega' \bar{\varepsilon}(\om-\om',\mathbf{r}) E_{\alpha}^{[\text{in}] }[\om',\mathbf{r}],\\
E_{\alpha}^{[\text{s}] }[\om,\mathbf{r}] &=-\omega^{2}\int\d\mathbf{r}'\int\d\omega'
G_{\alpha\beta}[\om,\mathbf{r}-\mathbf{r}']
\nn&\times
\bar{\varepsilon}[\om-\om',\mathbf{r}'] \,
E_{\beta}^{[\text{in}] }[\om',\mathbf{r}'],
\label{eri}
\end{align}
where $\bar{\varepsilon}[\om,\mathbf{r}]$ is the Fourier transform of $\bar{\varepsilon}(t,\mathbf{r})$ [cf.~(\ref{fou})],
and $G_{\alpha\beta}[\om,\mathbf{r}]$ is retarded Green's function of Helmholtz's operator [see Appendix \ref{helmo}]:
 \begin{align}
 &\{(\partial_\sigma\partial_\sigma+\om^2)\delta_{\alpha\beta}-\partial_\alpha\partial_\beta\}G_{\beta\gamma}[\om,\mathbf{r}]=\delta_{\alpha\gamma}\delta(\mathbf{r}),\\
 & G_{\alpha\beta}[\om>0,\mathbf{r}]=\Big(\delta_{\alpha\beta}+\frac{\partial_\alpha\partial_\beta}{\om^2}\Big) 
 G[\om,\mathbf{r}], ~r=|\mathbf{r}|,
 \label{sev} \\
 & G[\om,\mathbf{r}]=
\frac{e^{i\om r}}{(-4\pi) r},
\label{scal}
 \end{align}
where $ G[\om,\mathbf{r}]$ in (\ref{sev}) is the scalar Green's function.

Eq.~(\ref{eri}) can be related to the Doppler effect; see Appendix \ref{doppo}.
For ${E}_{\alpha}^{[\text{s}]  }[  \omega,\mathbf{r}]  $ we find from (\ref{eri}, \ref{Ert}):
\begin{align}
&{E}_{\alpha}^{[\text{s}]  }[\omega,\mathbf{r}]
=\frac{\om^2}{2\pi} \int d\mathbf{q}\,\sqrt{q}\,{\rm e}_{\beta\lambda}\left(  \mathbf{q}\right)  [V_{\alpha\beta}[
\mathbf{r,q},\omega-q]  
{a}_{\lambda}\left(  \mathbf{q}\right)  
\nn&\quad\quad\qquad
\label{10.a}%
+V_{\alpha\beta}[
\mathbf{r,-q},\omega+q]\, 
{a}_{\lambda}^{\dagger}\left(  \mathbf{q}\right)  ], \\
&V_{\alpha\beta}[ \mathbf{r,q},\omega-q] =\int d\mathbf{r}
^{\prime} 
G_{\alpha\beta}[  \omega,\mathbf{r-r}^{\prime}]
\bar{\varepsilon}[\om-q,\mathbf{r}']e^{i\mathbf{q\cdot r}'}. \nonumber
\end{align}
Note that for the time-independent dielectric, $\bar{\varepsilon}(\om)\propto \delta(\om)$, and only the annihilation operator containing part survives in (\ref{10.a}, \ref{11}), because (\ref{11}) refers to $\om>0$, and then $V_{\alpha\beta}[ \mathbf{r},-\mathbf{q},\omega+q]=0$ drops out. More generally, also the creation operator will contribute to (\ref{11}), which is the core of our effects. We now apply (\ref{10.a}, \ref{11}) to a space-time-modulated dielectric at rest. 

\section{Space-time-modulated dielectric}
\label{modulated}

\begin{figure}
\includegraphics[width=8.5cm]{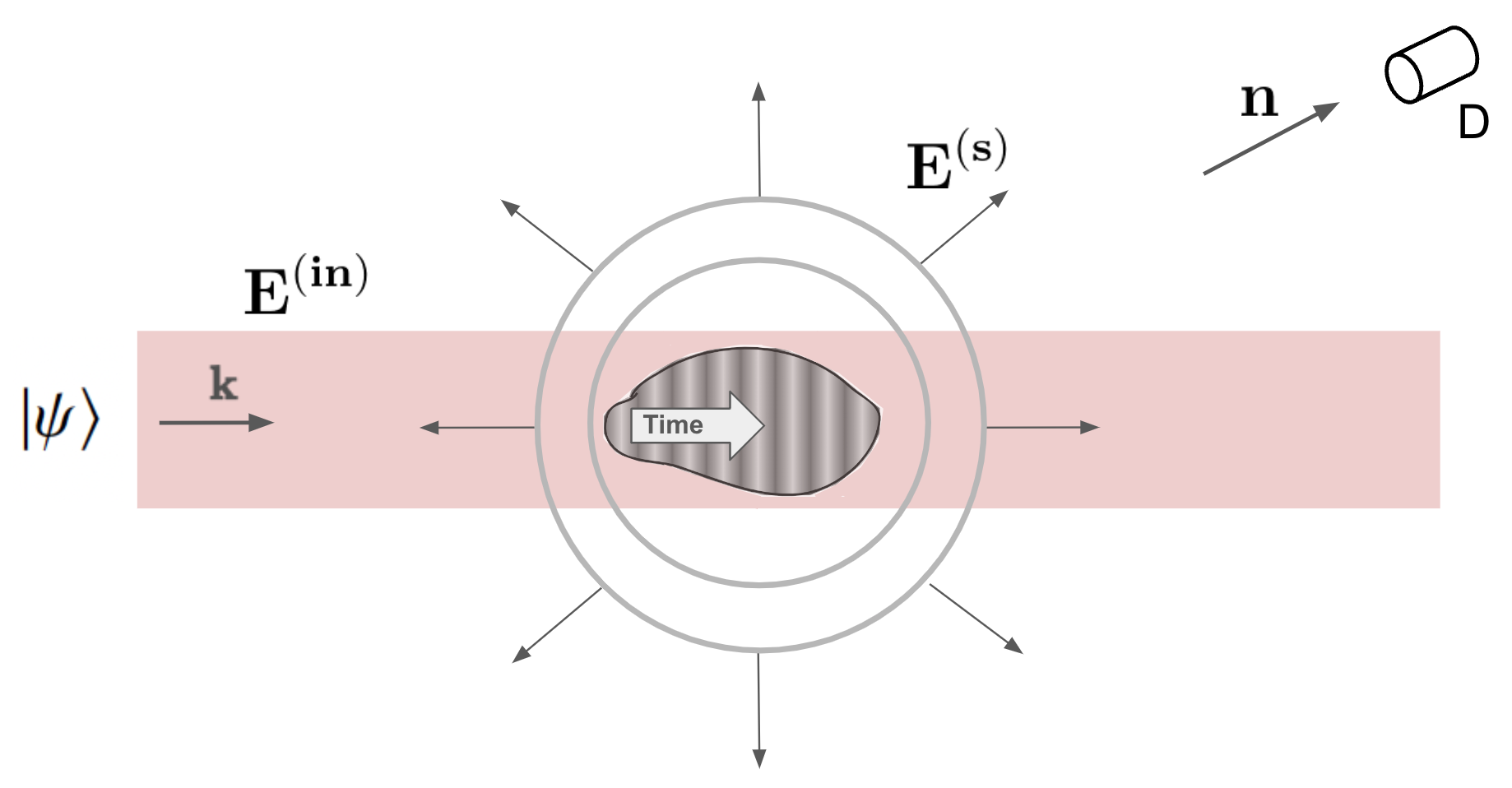}
\caption{Scattering from resting, space-time-modulated dielectrics (modulation devices are not shown). Shadowed domain denotes the incident field $\mathbf{E}^{[{\rm in}]}$. This can be a single-photon field with momentum $\mathbf{k}$ or the vacuum state of the field; cf.~(\ref{5.1}). $|\psi\rangle$ is the state of the field, which appears in the photodetection result (\ref{11}).
$\mathbf{n}$ is the unit vector towards the photodetector D that is placed in the far-field zone; see (\ref{halagaz}). In that zone, only the spherical scattered field $\mathbf{E}^{[{\rm s}]}$ is present if $|\mathbf{n\cdot k}|$ is sufficiently far from $|\mathbf{k}|$, i.e. no backscattering or forward scattering is detected. 
}
\label{fig1}
\end{figure} 

\subsection{ Vacuum response }
\label{vaco}

For simplicity, we work with the following model of space-time-modulated dielectric [see Fig.~\ref{fig1}]:
\begin{align}
\label{4.1}%
&\bar{\varepsilon}\left(  \mathbf{r,}t\right)  =\chi\left(  \mathbf{r}\right)
\left[  1+\eta\left(  x-ut\right)  \right], \\
& \bar{\varepsilon}[  \omega,\mathbf{r}]  =\chi(  \mathbf{r}%
)  \left(  \delta\left(  \omega\right)  +\frac{1}{u}\eta[
\omega/u]  e^{i\omega x/u}\right), \label{4.2}%
\end{align}
where $\chi\left(  \mathbf{r}\right)  $\ is a localized function that defines the overall shape of the dielectric, $\eta\left(x-ut\right)$ refers to the unknown structure along the $x$-axis that defines the modulation, and {$\eta[\om/u]$ is the Fourier transform of $\eta(x-ut)$ at the point $\om/u$} [cf.~(\ref{fou})]. The modulation speed $u>0$ can be larger than $c=1$ since it does not refer to the energy or information transfer \cite{pendry,caloz}. However, we shall not need $u>c=1$.

We calculate the vacuum response of the photodetector starting from (\ref{11}, \ref{10.a}, \ref{gabal})
for $|\psi\rangle=|0\rangle$: 
\begin{align}
I[\omega,\mathbf{r};|0\rangle\langle 0|]&=\frac{\om^4}{(2\pi)^2}\int\d \mathbf{q}\,q\,
\left(\delta_{\beta\gamma}-\frac{q_\beta q_\gamma}{q^2}\right)
\nn&\times
V^*_{\alpha\beta}[\mathbf{r},\mathbf{q},\omega+q]
V_{\alpha\gamma}[\mathbf{r},\mathbf{q},\omega+q],
\label{faso}
\end{align}
where only the negative frequency ($\om'<0$, i.e. creation operator containing) part of $\mathbf{E}^{\rm [s]}[\om',\mathbf{r}']$ contributes into (\ref{faso}). Employing (\ref{4.2}) in (\ref{faso}) we get more specifically
\begin{align}
I[\omega,\mathbf{r};|0\rangle\langle 0|]&=\frac{\om^4}{u^2(2\pi)^2}\int\d \mathbf{q}\,q\,
|\,\eta[{(\om+q)}/{u}]\,|^2 \nn&\times
\label{hope2}
\left(\delta_{\beta\gamma}-\frac{q_\beta q_\gamma}{q^2}\right)
\zeta_{\alpha\beta}(\mathbf{q,r})\zeta^*_{\alpha\gamma}(\mathbf{q,r}),\\ 
\label{hope3}
\zeta_{\alpha\beta}(\mathbf{q,r})=\int & \d\mathbf{r}'G_{\alpha\beta}[\om,\mathbf{r}-\mathbf{r}']\chi(\mathbf{r}')e^{
\frac{i(\om+q)x'}{u}-i\mathbf{q\cdot r}'
},
\end{align}
If $\chi(\mathbf{r}')$ is well-localized e.g. around $\mathbf{r}'\simeq 0$, (\ref{hope3}) can be simplified further assuming that the photodetector in (\ref{hope2}) is placed far away from the dielectric. In this far-field zone $|\mathbf{r}|\gg |\mathbf{r}'|$, the photodetector sees a spherical wave [see Appendix \ref{helmo} and recall $r=|\mathbf{r}|$],
\begin{align}
    G_{\alpha\beta}[\om,\mathbf{r-r}']\simeq(\delta_{\alpha\beta}-n_{\alpha} n_{\beta})\frac{e^{i\om r-i\om \mathbf{n\cdot r'}}}{(-4\pi)r}, ~
    \mathbf{n}=\frac{\mathbf{r}}{r}.
    \label{halagaz}
    \end{align}
Eq.~(\ref{hope2}) now simplifies, since (\ref{hope3}) reduces via (\ref{halagaz}) to the Fourier transform of $\chi(\mathbf{r})$:
\begin{align}
\label{ikos}
I[\omega,\mathbf{r};|0\rangle\langle 0|]&=
\frac{\pi^2\om^4}{u^2r^2}\int\d\mathbf{q}\,q\Big(1+\frac{(\mathbf{n\cdot q})^2}{q^2}\Big)
\\\times
\label{kikos}
&\Big|\eta\Big[ \frac{\om+q}{u} \Big] \,
\chi\Big[\mathbf{q}+\om\mathbf{n}-\mathbf{h}_x\frac{\om+q}{u}\Big]\Big|^2,\\
\chi[\mathbf{q}]&\equiv\int \frac{\d\mathbf{r}}{(2\pi)^3} e^{-i\mathbf{q\cdot r}}\chi(\mathbf{r}),
\label{ameoba}
\end{align}
where $\mathbf{h}_x=(1,0,0)$ is the unit vector of the $x$-axis. 

Note that for the finiteness of $I[\omega,\mathbf{r};|0\rangle\langle 0|]$ in (\ref{ikos}),
the product in (\ref{kikos}) should decay with $q$ sufficiently quickly.
For instance, if $\eta\left( x\right)  $ is Gaussian in
(\ref{4.1}), $\left\vert \eta\left[ \frac{\omega+q}{u}\right]  \right\vert ^{2}$ is
also Gaussian and the integral (\ref{hope2}) will be finite. 
The fact that (\ref{ikos}) can essentially depend on 
$\left\vert \eta\left[  \frac{\omega+q}{u}\right]  \right\vert ^{2}$ means that certain features of the susceptibility can be deduced from the vacuum response, i.e. without shining any light on the dielectric. 

\subsection{Example} 
\label{example3b}
In this subsection we recover $\hbar$ and $c$ to make the formulas more familiar.
For calculating (\ref{hope2}--\ref{halagaz}) we take the following models:
\begin{align}
    \chi(\mathbf{r})&=\chi\,\theta(a-|x|)\,\delta(y)\,\delta(z),
        \label{hag2} \\
   \eta(x-ut) &= \frac{\sin[(x-ut)/a]}{1+[(x-ut)/A]^2},
    \label{om}   
\end{align}
where $\theta(x)$ is the step-function, $\theta(x<0)=0$, $\theta(x>0)=1$,
$\chi$ has dimensionality ${\rm (length)}^2$ and has a meaning of an effective cross-section area, and 
$2a$ is the length. Eq.~(\ref{om}) indicates on harmonic modulation with broad switching driven by length-scale $A>0$, which ensures the proper behavior for large values of $(x-ut)$ in (\ref{om}). 
We assume a large $A$, i.e.,
\BEA
A\left[\frac{\om}{u}+\frac{1}{a}\right]\gg 1,
\EEA
and work out the Fourier transform (\ref{fou}) of (\ref{om})
[see Appendix \ref{evaluation}]: 
\BEA
\frac{1}{u^2}\left|\eta\left[-\frac{\omega+qc}{u}\right]\right|^2 
\simeq \frac{T}{4}
\delta\left(\om +cq-\frac{u}{a} \right),~~ T=\frac{A}{u},
\label{omni2}
\EEA
where $T$ relates to the coherence time of the modulation. 

We work out (\ref{ikos}--\ref{omni2}) with the final result:
\begin{align}
\label{vac} 
&I[\omega,\mathbf{r};|0\rangle\langle 0|]=
    \frac{\hbar \om^2}{c^3} \,\,
    \frac{ \om^2\pi^2 T \chi^2 }{(2\pi)^5 r^2 c a}f(b,n_x,\om),\\
&f(b,n_x,\om)= b^3 \,\theta \left(b\right)
\int^{1}_{-1}\d y\, \frac{\sin^2(by+\frac{a\om n_x}{c}-1)}{(by+\frac{a\om n_x}{c}-1)^2}
\nn&~\,\quad\qquad\quad \times
\Big (1+n_x^2 y^2+\frac{(1-y^2)(1-n_x^2)}{2}\Big) 
    \label{f} 
    \\
&b\equiv (u-\om a)/c.
\label{bb}
\end{align}
\comment{Here we employed the following calculation in spherical coordinates:
\BEA
&\int_0^\pi\d \theta F(\cos\theta) \sin\theta\int_0^{2\pi}\d \phi\times\nn 
&[1+(n_x\cos\theta +n_y\sin\theta \cos\phi
+n_z\sin\theta \sin\phi )^2],
\EEA
where $F(\cos\theta)$ is arbitrary. 
When we open the square and take the integral $\int_0^{2\pi}\d \phi $, we see that all cross-terms disappear. Hence
\BEA
&\int_0^\pi\d \theta F(\cos\theta) \sin\theta\int_0^{2\pi}\d \phi\times\nn 
&[1+n_x^2\cos^2\theta +n_y^2\sin^2\theta \cos^2\phi
+n_z^2\sin^2\theta \sin^2\phi ].
\EEA
Using 
\BEA
\int_0^{2\pi}\d \phi \cos^2\phi=\int_0^{2\pi}\d \phi \sin^2\phi=\pi,
\EEA
we find 
\BEA
&2\pi\int_0^\pi\d \theta F(\cos\theta) \sin\theta\times\nn
&[1+n_x^2\cos^2\theta +\frac{n_y^2+n_z^2}{2} \sin^2\theta ]\\
&=2\pi\int_{-1}^1\d y F(y) \times\nn
&[1+n_x^2y^2 +\frac{n_y^2+n_z^2}{2} (1-y^2) ]
\EEA
}
It is seen that (\ref{f}) contains the step function $\theta(b)$, which comes from (\ref{omni2}). This is a threshold effect: the existence of $I[\omega,\mathbf{r};|0\rangle\langle 0|]$ demands a sufficiently large $u$ in (\ref{bb}). 
We briefly mention the behavior of $f(b,n_x,\om)$ for each argument (the other two arguments held fixed):
$f(b,n_x,\om)$ maximizes at a specific value of $n_x$; $f(b,n_x,\om)$ is a growing function of $b>0$, and possibly a non-monotonous function of $\om$ (depending on fixed values of $n_x$ and $b$). 

Appendix \ref{evaluation} also calculates (\ref{vac}) for a realistic range of parameters and compares the result with the detection of a single-photon, as well as with the detection of laser light. The conclusion is that observing 
$I[\omega,\mathbf{r};|0\rangle\langle 0|]$ can be feasible because the corresponding photodetection signal can be larger than a photodetection signal from a single photon source; see Appendix \ref{evaluation}. 

Note that for $a\om/c\ll 1$, that is, when the wavelength of the detected photon (far field) is much larger than the dielectric size $a$, the dependence on $a$ disappears from the integral in (\ref{f}). However, overall $I[\omega,\mathbf{r};|0\rangle\langle 0|]$ in (\ref{vac}) still exhibits dependence on the dielectric size $a$.

\subsubsection{Relations with the anomalous Doppler effect} 
\label{ano}

When looking at (\ref{hope2}, \ref{omni2}) it is seen that a non-zero result is achieved only when 
the modulation speed $u$ is larger than $\om a$, where $\om$ is the frequency of the detected photon.
The difference $\frac{u}{a}-\om$ is the frequency $cq=c|\mathbf{q}|$ over which the integration 
in (\ref{hope2}) goes. Recall that this integration came from negative frequencies in (\ref{faso}). This 
suggests a relation with the anomalous Doppler effect, a mechanism by which an (internally) equilibrium body moving in a medium with a speed larger than the phase velocity of light in that medium can radiate \cite{ginzburg,ginzburg1996}. Cherenkov radiation is a limiting case of the anomalous Doppler effect \cite{tamm,boloto}. 

However, the analogy is limited in several respects: for (\ref{omni2}) the emitter is immersed in vacuum, and the modulation (not kinematic motion) speed should be larger than $\om a$, which is a geometric quantity, and not the phase velocity of light. Another difference is that (\ref{omni2}) is an approximate relation: it occurred due to the limit $A\to\infty$ in (\ref{om}). Our effect also exists without this limit, i.e. without the constraint $u>a\om$ implied by (\ref{omni2}). In contrast, Cherenkov radiation and the anomalous Doppler effect are strictly threshold-dependent effects \cite{ginzburg,boloto}. Also, there is a structural difference: Cherenkov radiation and the anomalous Doppler effect relate to cylindrical waves in contrast to (\ref{ikos}), which refers to spherical waves. 

\subsection{One-photon incident field} 
\label{shino}

We return to (\ref{10.a}--\ref{4.2}) and radiate the modulated dielectric with a single-photon state 
\begin{align}
\left\vert \psi\right\rangle =\int d\mathbf{q}\,C_{\lambda}\left(
\mathbf{q}\right)  \hat{a}_{\lambda}^{\dagger}\left(  \mathbf{q}\right)
\left\vert 0\right\rangle, ~ \int d\mathbf{q}|\,C_{\lambda}(  \mathbf{q})|^2=1, 
\label{5.1}%
\end{align}%
where $\left\vert 0\right\rangle $ is the vacuum state of the field, and the second relation in (\ref{5.1}) ensures $\left\langle \psi\vert\psi\right\rangle =1$. While the single-photon state was taken for clarity, similar results are obtained for a coherent state of the field that models laser light. 
{As the photodetector is placed in the far-field zone (see Fig.~\ref{fig1}), in (\ref{11}) we can take} $\mathbf{E}=\mathbf{E}^{[\text{s}] }$, i.e. only the scattered field contributes. Using (\ref{10.a}, \ref{11}) we find
\begin{align}
\label{30}
&I[\omega,\mathbf{r};|\psi\rangle\langle \psi|]=I[\omega,\mathbf{r};|0\rangle\langle 0|]+
\xi_\alpha\xi^*_\alpha,\\
&\xi_\alpha=
\om^2\int\frac{\sqrt{q}\,\d\mathbf{q}}{2\pi}C_{\lambda}\left(
\mathbf{q}\right)  V_{\alpha\beta}\left[  \mathbf{r},\mathbf{q},\omega-q\right]
e_{\beta\lambda}\left(  \mathbf{q}\right), \nonumber
\end{align}
where the vacuum contribution (\ref{hope2}, \ref{hope3}) enters additively, and where the non-vacuum contribution 
$\xi_\alpha\xi^*_\alpha$
comes from the positive frequencies. We now assume that the single-photon state (\ref{5.1}) has a well-defined momentum $\mathbf{k}$, which factorizes from the polarization:
\begin{align}
\label{gonch}
&C_{\lambda}\left(  \mathbf{q}\right)  =c_{\lambda}C\left( \mathbf{q}\right), ~C\left(  \mathbf{q}\right)~{\rm concentrates~at}~ \mathbf{q\simeq k},\\
&|c_1|^2+|c_2|^2=1.
\label{gonch2}
\end{align}
{Eq.~\eqref{30} confirms that, generally, all frequencies appear in $I[\omega,\mathbf{r};|\psi\rangle\langle \psi|]$.} This contrasts with the weak scattering of a non-modulated dielectric at rest, where $\mathbf{E}^{[{\rm s}]}$ contains only the incident frequency $|\mathbf{k}|$, hence the photodetector should be tuned to that frequency: $\om\simeq|\mathbf{k}|$. This fact is seen from (\ref{4.2}) after dropping out the modulation contribution. This fact is related to the classical Rayleigh limit \cite{we}. 

For clarity, we assume $|\mathbf{k}|\not=\om$ in (\ref{30}), i.e. the detection frequency differs from the incident frequency. Together with (\ref{gonch}), this allows us to exclude the term $\propto \delta(\om)$ in (\ref{4.2}). We find from (\ref{30}, \ref{gonch}, \ref{gonch2}, \ref{10.a}): 
\begin{align}
I[\omega,&\mathbf{r};|\psi\rangle\langle \psi|]-I[\omega,\mathbf{r};|0\rangle\langle 0|]=
\frac{\pi^2\om^4\sigma}{u^2r^2}
\nn&\times
\label{mikado}
\Big|\eta\Big[ \frac{\om-k}{u} \Big] \,
\chi\Big[\mathbf{k}-\om\mathbf{n}+\mathbf{h}_x\frac{\om-k}{u}\Big]\Big|^2,\\
\sigma =& \Big(1  -  |c_\lambda\mathbf{e_\lambda(k)\cdot n}|^2\Big)
\left\vert \int\d\mathbf{q}\,\sqrt{q}\,
C\left(  \mathbf{q}%
\right)  \right\vert ^{2},
\label{yunus}
\end{align}
where $\mathbf{h}_x=(1,0,0)$ and $\chi[\mathbf{q}]$ is defined in (\ref{ameoba}). Eq.~(\ref{yunus}) 
relates to the scalar product of the initial polarization vector $c_\lambda\mathbf{e_\lambda(k)}$ with the observation direction $\mathbf{n}=\mathbf{r}/|\mathbf{r}|$.

The factor $|\eta[ {(\om-k)}/{u} ]|^2$ in (\ref{mikado}) can violate the Rayleigh limit, e.g. for $\om=k/2$ and a sufficiently small $u$. The validity of this limit implies that the far-field photodetection response contains only $|\eta[\om']|^2$ with $\om'\simeq k$ \cite{we}. A similar violation of Rayleigh limit takes place on the level of 
$\chi[...]$ in (\ref{mikado}). Note that it implies an enhanced detection of the dielectric inhomogeneity along the $x$-axis, i.e. along the direction of modulation. 

There are examples, where Rayleigh limit is violated for evanescent fields \cite{nano}, but we are not aware of far-field violations of this limit besides the effects related to specific quantum correlation (entanglement) features of the incident light \cite{Shih,Abouraddy,Dowling,monken,liu,we}. Here no such correlated states are needed, because violations of the classical limit occur due to the space-time modulation.  

\section{Moving dielectric} 
\label{moving}

\subsection{Scattering within relativistic electrodynamics} 

\begin{figure}
\includegraphics[width=6cm]{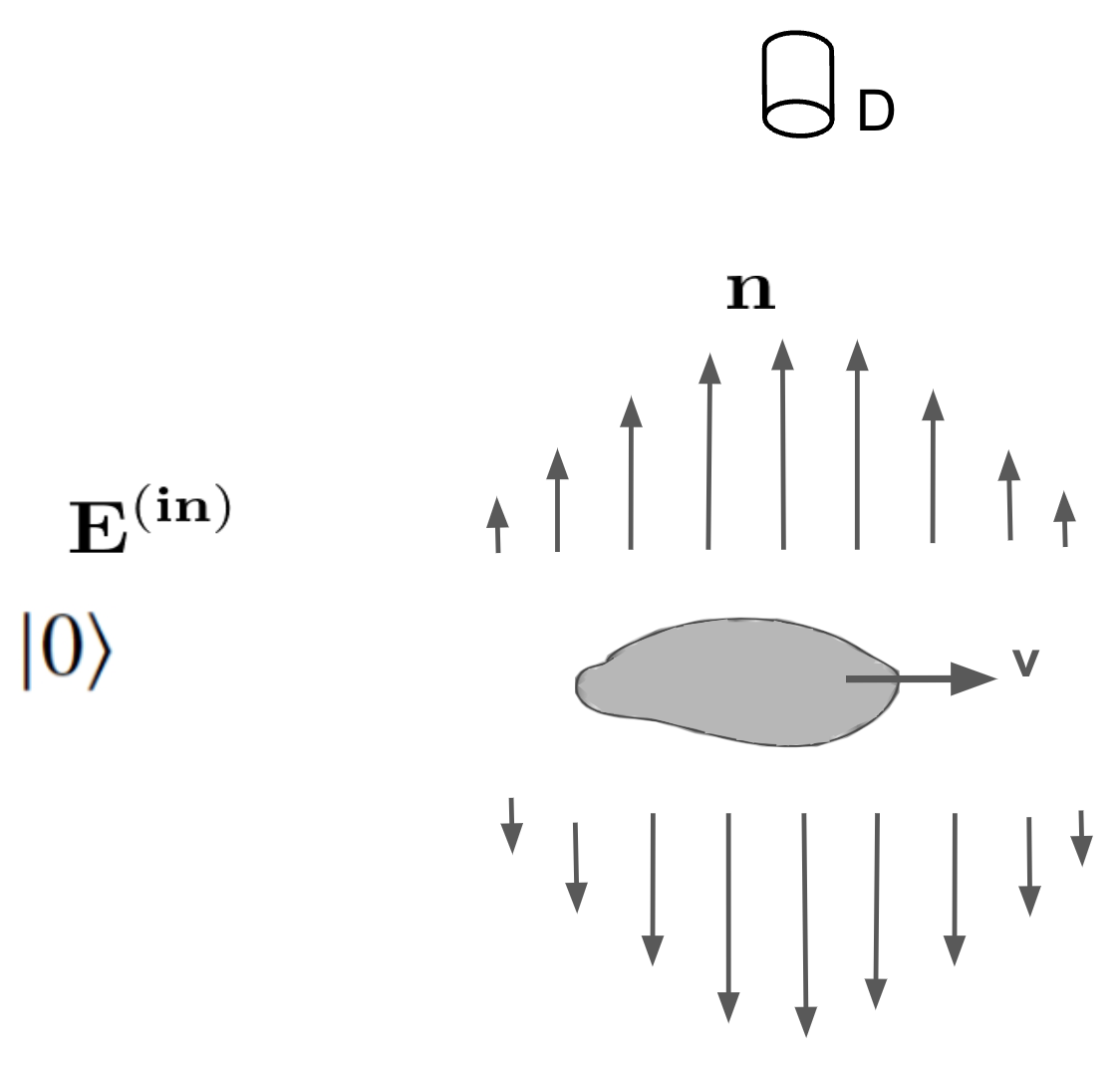}
\caption{Vacuum scattering from dielectric moving with speed $v$. The incident field $\mathbf{E}^{[\text{in}]  }$ is an operator; cf.~(\ref{6}, \ref{Ert}). The scattered field is cylindrical; see Fig.~\ref{fig1} for other notations. 
}
\label{fig2}
\end{figure}

Consider a (non-magnetic) dielectric which in the laboratory frame moves along $x$-axis with a constant velocity
$v$; see Fig.~\ref{fig2}. We study this problem within the framework of relativistic electrodynamics \cite{hermann,landau2}, which will allow a consistent treatment of space-time-dispersion. Relativistic consideration was not needed for the modulated situation (\ref{4.1}, \ref{4.2}), because the modulation speed (not limited by special relativity, since there is no energy transfer or information transfer) can be comparable to or even greater than the speed of light \cite{pendry,caloz}. 

\comment{and $c=\hbar=1$. Latin indices assume values $0,1,2,3$ and refer to 4-vectors [cf.~(\ref{a2})]:}

We set the following standard notations:
\begin{align}
&x^i=(t,\mathbf{r}),~~~\mathbf{r}=(x^1,x^2,x^3)=(x,y,z),\\
&A^i=(A^0,\mathbf{A}), ~~~A_i=g_{ik}A^k=(A_0,-\mathbf{A}),\nn
& g_{ik}={\rm diag}[1,-1,-1,-1],\nn
&\mathbf{A}=(A^1,A^2,A^3)=(A_x,A_y,A_z).
\end{align}
The electromagnetic field tensor 
\BEA
F^{ik} =\partial^iA^k-\partial^kA^i,
\EEA
where $A^i$ is the 4-vector-potential, satisfies the following
wave-equation \cite{hermann,pal} (see Appendix \ref{minko}, Eqs.~(\ref{8}, \ref{88}))
\begin{align}
-\partial_{i}F^{ik}(  x)  &=\partial_{i}\int \d^4\hat x\, \bar{\varepsilon}(  x,\hat x)  
\nn&\times 
(F^{is}u_{s}u^{k}-F^{ks}\,u_{s}\,u^{i})(\hat x),\label{drug}\\
u^{i}&=(  \gamma,\gamma v,0,0),\quad \gamma=(1-v^2)^{-1/2},
\end{align}
where $u^i$ is the 4-vector of velocity {of the dielectric} \cite{landau2}, and $\bar{\varepsilon}=\varepsilon-1$ is the dielectric susceptibility. {Eqs.~(\ref{8}, \ref{88}) were deduced for a dielectric without space-time dispersion, but its generalization towards (\ref{drug}) is straightforward, as it respects the relativistic covariance. Note that $\bar{\varepsilon}(  x,\hat x)$ in (\ref{drug}) contains both inhomogeneity in space, and space-time dispersion.} 

Recall that $F^{ik}$ is gauge-invariant, but contains redundant variables. 
In the Lorenz gauge $\partial_iA^i=0$, we get from (\ref{drug})
\BEA
&-\partial_{i}\partial^{i}A^{k}=
\partial_{i}\int \d\hat x\, \bar{\varepsilon}(  x,\hat x)\,  \{\,
u^{s}u^{k}( \partial^{i} A_{s} -\partial_{s} A^{i} ) \nonumber\\
& -u^{s}u^{i}(  \partial ^{k} A_{s} -\partial_{s} A^{k})  \,\} (\hat x).
\label{444}
\EEA
The virtue of the Lorenz gauge is that (\ref{444}) is an inhomogeneous wave-equation. 

Now we need to determine the form of $\bar{\varepsilon}(  x,\hat x)$ in (\ref{444}). Recall that this equation is written in the 
laboratory frame, where the dielectric moves with velocity $v$. We need to look at the rest frame of the dielectric with coordinates $x'^{i}$. Our first assumption is that in the rest frame the space inhomogeneity separates from the space-time dispersion: 
\begin{align}
\label{L}
&\bar{\varepsilon}(  x',\hat x')=\bar\varepsilon[x'^i-\hat x'^i; x'^i],\\
& \bar\varepsilon[x'^i-\hat x'^i; x'^i] =0 ~~{\rm for}~~
(x'^i-\hat x'^i)(x'_i-\hat x'_i)\leq 0,
\label{dodosh}
\end{align}
where (\ref{dodosh}) refers to causality. {Next, we assume that the inhomogeneity in (\ref{L}) refers to the space only, while the dispersion refers only to the time (i.e. the proper space-dispersion is absent):  
$\bar\varepsilon[x'^i-\hat x'^i; x'^i]=\bar\varepsilon[t'-\hat t'; \mathbf{r} ']$. Now in the laboratory frame $\bar{\varepsilon}(x,\hat x)$ in (\ref{drug}) is obtained from (\ref{L}) after making in (\ref{L}) the Lorentz transformation}
\BEA
&x'=\gamma(x-vt),~~t'=\gamma(t-vx),~~y'=y,~~z=z'.
\label{LT}
\EEA

Within the first-order Born approach to scattering [cf.~(\ref{6})], on the left-hand-side of (\ref{444}), we should change $A^k\to A^{{\rm [s]}\,k}$, while $A^k\to A^{{\rm [in]}\,k}$ in the right-hand-side. Now $A^{{\rm [in]}\,k}$ is given via (\ref{Ert}), where $A^{{\rm [in]}\,0}=0$ and $\mathbf{E}^{{\rm [in]}}=-\partial_t{\mathbf{A}}^{{\rm [in]}}$. Hence we transform (\ref{drug}--\ref{444}) as: 
\begin{align}
&(\Delta-\partial^2_t)A^{{\rm [s]}\, k}(\mathbf{r},t)\nn& = 
\frac{\gamma u^k}{2\pi}\int\d \mathbf{q}\,
\frac{C^i_\lambda(\mathbf{q})}{q-vq^1}
\{a_\lambda(\mathbf{q})\,e^{-iq_lx^l}\,\partial_i\hat\varepsilon[\mathbf{q},\mathbf{r},t]+{\rm h.c.}\}\nn&+
\label{natrium}
\frac{\gamma^2}{2i\pi}\int\d \mathbf{q}\,
C^k_\lambda(\mathbf{q})
\{a_\lambda(\mathbf{q})\,e^{-iq_lx^l}\,\hat\varepsilon[\mathbf{q},\mathbf{r},t]
-{\rm h.c.}\},\\
&C^k_\lambda(\mathbf{q})=q^{-1/2}(q-vq^1)\nn
&\,\,\,\,\quad\quad\times
[v\,{\rm e}^1_\lambda(\mathbf{q})\, q^k+{\rm e}^k_\lambda(\mathbf{q})(q-vq^1)],
\end{align}
where $q^l=(|\mathbf{q}|,\mathbf{q})$, $q_lq^l=0$, and where we denoted
\BEA
{\rm e}^k_\lambda(\mathbf{q})=(0,\mathbf{e}_\lambda(\mathbf{q})).
\EEA
In (\ref{natrium}), $\hat\varepsilon[\mathbf{q},\mathbf{r},t]$ is defined as the Fourier-transform over its space-time dispersion features, while its space-time inhomogeneity is left intact:
\begin{align}
\label{sodium}
\hat\varepsilon[\mathbf{q},\mathbf{r},t]&=\int\d^4\hat x\, e^{i\hat q_l\hat x^l} 
\bar\varepsilon[\hat x; \gamma(x-vt),y,z], \\ 
\hat q^l &=(\gamma(q-vq^1), \gamma(q^1-vq),q^2,q^3).~
\label{sodium2}
\end{align}
Note from (\ref{sodium}) that for $\hat\varepsilon[\mathbf{q},\mathbf{r},t]$ we have 
$u^i\partial_i \hat\varepsilon[\mathbf{q},\mathbf{r},t]=0$.

Eq.~(\ref{natrium}) shows that the scattered field $A^{{\rm [s]}\, k}(\mathbf{x},t)$ depends also on space-derivatives of $\bar\varepsilon$. Such a dependence is absent in the phenomenological treatment; cf.~(\ref{eri}).
We note that $\hat\varepsilon$ in (\ref{sodium}) is Fourier-transform of the dispersive part of $\bar\varepsilon$; i.e. (\ref{sodium}) refers to the usual form of the dispersive dielectric susceptibility \cite{landau}. 

For simplicity, we no longer consider space-dispersion, i.e. $\hat\varepsilon$ in (\ref{sodium}) reads
\BEA
\label{kobzar}
\hat\varepsilon[\gamma(q-vq^1), \gamma(x-vt),y,z],
\EEA
where $\gamma(q-vq^1)$ is the frequency of the dispersive susceptibility. 
Next simplification occurs when we consider only the transversal component of the scattered field, i.e. 
$A^{{\rm [s]}\, 2}(\mathbf{x},t)=A^{{\rm [s]}}_y(\mathbf{x},t)$ or $A^{{\rm [s]}\, 3}(\mathbf{x},t)$. Then only (\ref{natrium}) survives; cf.~(\ref{drug}). Note that within the Lorenz gauge, the separate components of the vector potential $A^i$ are meaningful and measurable \cite{allahverdyan2016electromagnetic}.

Thus Fourier-transforming $A^{{\rm [s]}}_y(\mathbf{x},t)$ and working out as in (\ref{fou}, \ref{a11}--\ref{eri}) we get
\begin{align}
&A^{{\rm [s]}}_y[\om,\mathbf{r}] = \frac{\gamma}{2iv\pi}\int\d \mathbf{q}\,
C_{y\,\lambda}(\mathbf{q})\nn
&\times
\{a_\lambda(\mathbf{q})\,
{\cal V}[\mathbf{q},\gamma(q-vq_x),\om-q,\mathbf{r}]
\nn&-
a^\dagger_\lambda(\mathbf{q})\,{\cal V}[-\mathbf{q},-\gamma(q-vq_x),\om+q,\mathbf{r}]\},
\label{nush}\\
&{\cal V}[\mathbf{q},\gamma(q-vq_x),\om-q, \mathbf{r}] =
\int\d \mathbf{r}'\,G[\om,\mathbf{r}-\mathbf{r}']\nn
&\times
e^{i\mathbf{q}\mathbf{r}'+\frac{i(\om-q)x'}{v} }\, \, \widetilde\varepsilon[\gamma(q-vq_x),-\frac{\om-q}{\gamma v},y',z'],
\label{anush}
\end{align}
where $G[\om,\mathbf{r}-\mathbf{r}']$ is defined in (\ref{scal}),
and where $\widetilde\varepsilon$ is the Fourier-transform of (\ref{kobzar}):
\begin{align}
\widetilde\varepsilon[\gamma(q-vq_x),-\frac{\om-q}{\gamma v},y,z]&=
\int\frac{\d x}{2\pi}e^{-\frac{ix(\om-q)}{\gamma v}}\nn&\times 
\hat\varepsilon[\gamma(q-vq_x), x,y,z].
\label{wilson}
\end{align}
Structurally, (\ref{nush}) is similar to (\ref{10.a}). Several differences relate to polarization factors (plus the difference in Green's functions), but the major difference will be in the far-field radiation, which is absent in (\ref{nush}); see below. 

\comment{To evaluate (\ref{z1}, \ref{z2}) we adopt the following model for $\bar\ep(x^1,x^2,x^3)$
\begin{align}
\label{model}
\bar\ep(x^1,x^2,x^3)&=\bar\ep(x^1) \delta(x^2)\delta(x^3),\\
\bar\ep(\om,x^1,x^2,x^3)&=\frac{1}{\gamma v}e^{i\om x^1/v}\bar\ep\Big[ \frac{\om}{\gamma v}\Big] \delta(x^2)\delta(x^3),
\end{align}
where $\bar\ep\Big[ \frac{\om}{\gamma v}\Big]$ is the space-Fourier transform along $x^1$-axis; see (\ref{fou-1}, \ref{fou-2}). 
Eq.~(\ref{model}) leads to [see section \ref{bessel} for details]:
\begin{align}
\nu(\by,-\bq,\om &+q)=\frac{1}{(-4\pi)\gamma v}e^{iy^1(\frac{\om+q}{v}-q^1)}
\bar\ep\Big[ \frac{\om+q}{\gamma v} \Big]
\nn&\quad\times
\label{bull}
\int\d x\,\frac{e^{i\om\sqrt{x^2+\rho^2} +ix(\frac{\om+q}{v}-q^1) }}{\sqrt{x^2+\rho^2}}\\
&= \frac{1}{(-2\pi)\gamma v}e^{iy^1(\frac{\om+q}{v}-q^1)}
\bar\ep\Big[ \frac{\om+q}{\gamma v} \Big]
\nn&\quad\times 
K_0\left(\rho\sqrt{\Big(\frac{\om+q}{v}-q^1\Big)^2-\om^2}\right),
\label{karachi}
\\
\rho&=\sqrt{(y^1)^2+(y^2)^2},
\label{bu}
\end{align}
where $K_0[.]$ is the corresponding Bessel's function. Eq.~(\ref{karachi}) extendens for 
$\nu(\by,\bq,\om-q)$, but note that in (\ref{karachi}) we got $\sqrt{\Big(\frac{\om+q}{v}-q^1\Big)^2-\om^2}>0$, since $v<1$.
}

\subsection{Vacuum response} 

Eqs.~(\ref{nush}, \ref{anush}) imply for the vacuum response 
\begin{align}
\widetilde I[\omega,\mathbf{r};&|0\rangle\langle 0|]=\langle 0|A^{{\rm [s]}\dagger}_y [\om,\mathbf{r}]
A^{{\rm [s]}}_y[\om,\mathbf{r}]|0\rangle=\frac{\gamma^2}{4\pi^{2}v^2}
\nn&\times
\int\d\mathbf{q}\, q^{-1}(q-vq_x)^2 ((q-vq_x)^2-[1-v^2]q_y^2) 
\nn&\times 
\Big|{\cal V}[-\mathbf{q},-\gamma(q-vq_x),\om+q,\mathbf{r}]\Big|^2,
\label{vv}
\end{align}
where we employed (\ref{gabal}) for working out $C_{y\lambda}C_{y\lambda}$. A positive vacuum-response (\ref{vv}) is determined by the negative-frequency part of (\ref{nush}); cf.~(\ref{faso}). The vacuum effect is a consequence of the relative motion of the detector and scatterer, and it does not contradict the principle of relativity. 

\comment{The energy of the photodetector response is taken from the kinetic energy of the moving dielectric, but the considered theory is not subtle enough to account for those energy losses. }

To simplify (\ref{vv}), we shall work with a simpler model of a thin rod (cf.~ Sec. \ref{example3b})
\begin{align}
\label{model3}
&\hat\varepsilon[\hat\om;x,y,z]=\hat\varepsilon[\hat\om; x] \delta(\boldsymbol{\rho}), 
&\boldsymbol{\rho}=(y,z), ~~ \rho=|\boldsymbol{\rho}|.
\end{align}
We now employ (\ref{model3}) in (\ref{wilson}) that leads to simplifying (\ref{anush}) due to $\delta(\boldsymbol{\rho})$. 
Now in (\ref{vv}) we go to spherical coordinates, i.e., replace $(q_x, q_y, q_z)$ by $q_x=q\cos\theta$, $q_y=q\sin\theta\cos\phi$, $q_z=q\sin\theta\sin\phi$, where $q=\sqrt{q_x^2+q_y^2+q_z^2}\geq 0$, $\theta\in [0,\pi]$, and $\phi\in [0,2\pi]$. Then, we take the integral over $\phi$, and introduce new integration variables $\xi=\cos\theta$ and $q'=q(1-v\xi)/\om$. Eventually, we find (denoting $q'$ by $q$ again): 
\comment{Eq.(\ref{hopon0}) is deduced from (\ref{sandy}). Eq.(\ref{hopon}) is deduced from (\ref{hopon0}. )
\BEA
& \widetilde I[\omega,\rho;|0\rangle\langle 0|]=\frac{\gamma^2}{v^2(2\pi)^4}\int\d \mathbf{q}\,
\Big|\widetilde\varepsilon\Big[\gamma(q-vq_x);\frac{\om+q}{v\gamma}\Big]\Big|^2 \times\nn
&|K_0\left(\rho\sqrt{b^2-\om^2}\,\right)|^2 \, q^{-1}\,(q-vq_x)^2\times\nn 
&((q-vq_x)^2-[1-v^2]q_y^2),
\label{sandy}
\EEA
\BEA
\label{hopon0}
& \widetilde I[\omega,\rho;|0\rangle\langle 0|]=\frac{\gamma^2}{v^2(2\pi)^3}\int_0^\infty\d q\,q^5 \int_{-1}^{1}\d \xi\nn
&\Big|\widetilde\varepsilon\Big[\gamma q(1-v\xi);\frac{\om+q}{v\gamma}\Big]\Big|^2 \times\nn
&K^2_0\left(\frac{\rho}{v}\sqrt{[\om+q(1-v\xi)]^2-\om^2v^2}\,\right) \times\nn 
&(1-v\xi)^2[\,(1-v\xi)^2-\frac{(1-v^2)(1-\xi^2)}{2}\,]
\EEA}
\begin{align}
\label{hopon}
\widetilde I[\omega,\rho;|0\rangle\langle 0|] &= \frac{\om^6\gamma^2}{v^2(2\pi)^3}\int_0^\infty\d q\,q^5 \int_{-1}^{1}\d \xi(1-v\xi)^{-2}
\nn&\times
\left|\widetilde\varepsilon\Big[\gamma q\om;\frac{\om(1-v\xi+q)}{v\gamma(1-v\xi)}\Big]\right|^2 
\nn&\times
\left[\,1-\frac{(1-v^2)(1-\xi^2)}{2(1-v\xi)^2}\,\right]
\nn&\times
K^2_0\left(\frac{\rho\om}{v}\sqrt{[1+q]^2-v^2}\,\right),
\end{align}
where $\widetilde\varepsilon\Big[\gamma q\om;\frac{\om(1-v\xi+q)}{v\gamma(1-v\xi)}\Big]$ is defined via (\ref{wilson}), where $(y,z)$ is omitted, and where $K_0$ is Bessel's $K_0$-function \cite{bessel}. Its argument $\frac{\rho\om}{v} \sqrt{[1+q]^2-v^2}$ is non-negative. Since $\widetilde I$ in (\ref{hopon}) depends on $\rho$ only, it refers to cylindrical fields.
This is due to the thin-rod model (\ref{model3}) and the anisotropy introduced by the motion along the $x$-axis. The spherical far-field limit (\ref{halagaz}) does not apply to the moving dielectric e.g. because $|\widetilde{\varepsilon}|^2$ in (\ref{hopon}) does not depend on $x$, and hence does not nullify for $x\to\pm\infty$. The known asymptotics 
\begin{align}
K_\nu(x\gg 1)\simeq e^{-x} \, \sqrt{{\pi}/{(2x)}},
\label{asym}    
\end{align}
means that the integral $\int_0^{\infty}\d q$ in (\ref{hopon}) is finite.  
Indeed, $|\widetilde\varepsilon|^2$ in (\ref{hopon}) is generally such that due to the temporal dispersion (i.e. its dependence on $q-vq_x$), $|\widetilde\varepsilon|^2\to 0$ for $q\to \infty$ \cite{landau}. However, this tendency cannot compensate for the factor $q^5$, i.e. (\ref{asym}) is needed for the finiteness of $\int_0^{\infty}\d q$ in (\ref{hopon}). 

\comment{
In particular, (\ref{hopon}) applies to point particles. 
This situation differs from the modulated case (\ref{hope2}) which did not apply to point particles, where
$|\eta[{(\om+q)}/{u}]|^2={\rm const}$ (the interior of such a particle cannot be modulated). 
}
\comment{we assume the Debye model \cite{landau}:
\BEA
& \Big|\widetilde\varepsilon\Big[\gamma q\om;
\Omega
\Big]\Big|^2 =\frac{\bar\varepsilon_{\rm s}^2\sin^2[\frac{\Omega L}{2}]}{\pi^2\Omega^2(1+\tau^2\gamma^2\om^2q^2)^2}
\nn
&\Omega\equiv \frac{\om(1-v\xi+q)}{v\gamma(1-v\xi)}>0,
\label{model}
\EEA
where $\bar\varepsilon_{\rm s}$ is the static dielectric susceptibility, and $\tau$ is the inverse frequency parameter of the Debye model \cite{landau}. 
}

\subsection{Example: point-like dielectric}

{
We focus on the following simple example in (\ref{hopon}, \ref{model3}): 
\begin{align}
    \hat\varepsilon[\hat\om,x]= \lambda\delta(x)\,   \hat\varepsilon[\hat\om], 
\label{alis}
\end{align}
where $\lambda$ is a dimensional constant. Eqs.~(\ref{model3}, \ref{alis}) mean that we consider a point-like object.

Now the integral over $\xi$ in (\ref{hopon}) can be taken, and we get
\BEA
&&\widetilde I[\omega,\rho;|0\rangle\langle 0|] = \frac{\om^6\gamma^4\,\phi(v)\lambda^2}{(2\pi)^3}\int_0^\infty\d q\,q^5 
\left|\widetilde\varepsilon\Big[\gamma q\om\Big]\right|^2 
\nn \label{nono}
&&\times K^2_0\left(\frac{\rho\om}{v}\sqrt{[1+q]^2-v^2}\,\right),\\
&&\phi(v)=\frac{2}{v^5}(v-v^3(1-v^2)-(1-v^2)^2\,{\rm arctanh}[v]).~~~~~
\EEA }
We turn to the ultrarelativistic limit $v\to 1$ (i.e. $\gamma\to\infty$) in (\ref{nono}). 
For $|\hat\varepsilon(\gamma q\om)|^2$ in (\ref{nono}) we note the following relation that is valid for all materials in
the considered large-frequency limit \cite{landau}:
\BEA
\label{asy}
|\hat\varepsilon[\gamma q\om]\,|^2
\underset{\gamma\to \infty}{\mapsto}\om_{\rm e}^4 (\gamma\om q)^{-4},~~ 
\om_{\rm e}=\sqrt{4\pi N_{\rm e} {\rm e}^2/m_{\rm e} },
\EEA
where $\omega_e$ is the plasma frequency with  $N_{\rm e}$ being the electron density in the material; ${\rm e}$ and $m_{\rm e}$ are (resp.) the electron charge and mass. Note that $\gamma^{-4}$ in (\ref{asy}) matches to $\gamma^{4}$ in $\frac{\om^4\gamma^4}{8\pi^5}$; see (\ref{nono}). The origin of (\ref{asy}) is that in the high-frequency limit, the material reacts to the electric field via its (free) electrons. Note that (\ref{asy}) is naturally incorporated into the Drude model. 

We get from (\ref{nono}, \ref{asy}):
\BEA
\label{nan}
&&\widetilde I[\omega,\rho;|0\rangle\langle 0|] = \frac{2\om^2\om_{\rm e}^2\lambda^2}{(2\pi)^3}\,J(\rho\om), \\
&&J(\rho\om)=\int_0^\infty\d q\,q 
K^2_0\left(\rho\om\sqrt{[1+q]^2-1}\,\right)\nonumber \\
&&=\int_0^\infty\d u\,\frac{u(\sqrt{u^2+1}-1)}{\sqrt{u^2+1}} 
K^2_0\left(\rho\om u\right).
\label{nano}
\EEA
To present the asymptotic behavior of $J(\rho\om)$ for $\rho\om\gg 1$ and $\rho\om\ll 1$
we write it as follows
\BEA
\label{gaza1}
&&J(\rho\om)=c_1(\rho\om) (\rho\om)^{-4},\\
&&J(\rho\om)=c_2(\rho\om) (\rho\om)^{-2},
\label{gaza2}
\EEA
where $c_1(\rho\om)$ [$c_2(\rho\om)$] is a slowly varying function for $\rho\om\gg 1$ [$\rho\om\ll 1$]. Now $c_1(1)=0.076$,
$c_1(10)=0.163$ and $c_1(\rho\om>20)\to 0.166$. Likewise, $c_2(\rho\om)$ tends to $0.5$ for $\rho\om\to 0$, e.g., 
$c_1(0.1)=0.35$, $c_1(0.01)=0.478$, {\it etc}. 

{
Hence, (\ref{gaza1}) [(\ref{gaza2})] describes $J(\rho\om)$ at large [small] values of $\rho\om$. The asymptotic behavior $(\rho\om)^{-4}$ in (\ref{gaza1}) can be obtained by employing (\ref{asym}) in (\ref{nano}) and expanding $\frac{u(\sqrt{u^2+1}-1)}{\sqrt{u^2+1}} \simeq \frac{1}{2}u^3$ for small $u$, which is the dominant integration domain. Likewise, $(\rho\om)^{-2}$ in (\ref{gaza2}) can be obtained by changing in (\ref{nano}) variables $u'=\rho\om u$, and noting that for a small $\rho\om$ we have $\rho\om\sim u'$ in the dominant integration domain. }

Eqs.~(\ref{gaza1}, \ref{gaza2}) show that in the ultrarelativistic limit, the detection of the vacuum signal is a long-range effect, though it is not a far-field effect, because for cylindrical waves the far-field means intensity $\propto\rho^{-1}$. 

Finally, let us return to (\ref{nono}, \ref{asym}) and note that outside the ultra-relativistic limit, the behavior of $\widetilde I[\omega,\rho;|0\rangle\langle 0|]$ in (\ref{nono}) is (roughly) an exponential function of $\rho\om$:
\BEA
\label{andersen}
\widetilde I[\omega,\rho;|0\rangle\langle 0|]
={\cal F}(\rho\om)\, e^{-2\frac{\rho\om}{v}\sqrt{1-v^2}},
\EEA
where ${\cal F}(\rho\om)$ is not an exponential function of its argument. Eq.~(\ref{andersen}) is found from using 
(\ref{asym}) in (\ref{nono}), changing the variable $w=\sqrt{(1+q)^2-v^2}$, and noting that $w\in[\sqrt{1-v^2},\infty]$. 
Eq.~(\ref{andersen}) also describes the behavior of $\widetilde I[\omega,\rho;|0\rangle\langle 0|]$ for $v\to 0$.

\comment{
The previous treatment of this situation contained an error: $\sin^2[\Omega a]$ was taken to be 1/2 in the ultra-relativistic limit, which was incorrect. In fact the correct treatment shows that $\sin^2[\Omega a]$ has to be retained till the end. 
12.2024 The treatment of this thin rod example shows that in the ultra-relativistic situation the effect is long-range ($\rho^{-2}$, but contains a weak dependence on $\gamma$ such that it goes to zero for $v\to 1$. Below only the correct formulas are retained.

\subsection{Example: thin rod}
We assume in (\ref{model3}) a fixed linear size $2a$;
cf.~(\ref{hag2})
\footnote{The sharp boundary assumption made in (\ref{alisa}) will have to be revised, if more general quantities in (\ref{natrium}) (which involve space-derivatives of $\widetilde\varepsilon$) are to be calculated.  }:
\begin{align}
    \hat\varepsilon[\hat\om,x]= \theta(a-|x|)\,   \hat\varepsilon[\hat\om], 
\label{alisa}
\end{align}
where $\theta(x)$ is the step-function.
Eq.~(\ref{hopon}) now reads
\begin{align}
\widetilde I[\omega,\rho;|0\rangle\langle 0|] &= \frac{\om^4\gamma^4}{8\pi^5}
\int_0^\infty\d q\,q^5 \int_{-1}^{1}\d \xi(1-v\xi+q)^{-2}
\nn&\times
\label{cc1}
|\hat\varepsilon[\gamma q\om]\,|^2
\, K^2_0\left(\frac{\rho\om}{v}\sqrt{[1+q]^2-v^2}\,\right) 
\\&\times
\sin^2[\Omega a] \left[\,1-\frac{(1-v^2)(1-\xi^2)}{2(1-v\xi)^2}\,\right],
\label{cc2}\\
\Omega & \equiv \frac{\om(1-v\xi+q)}{v\gamma(1-v\xi)}>0.
\end{align}
In the ultra-relativistic limit $v\to 1$, $\gamma\to \infty$, we can employ
for the second term in (\ref{cc2}):
\BEA
\label{cc5}
(1-v^2)\int_{-1}^1\d \xi\,\frac{(1-\xi^2)}{(1-v\xi+q)^2(1-v\xi)^2}\underset{v\to 1}{\longmapsto} 0.
\EEA
}

\section{Summary}

{
The question asked in this paper is whether rectilinear motion, or space-time modulation of dielectric susceptibility, can lead to new mechanisms of dielectric structure determination. These two problems are studied together because the space-time modulation makes the dielectric susceptibility of a resting dielectric formally similar to that of a moving dielectric. 
The literature suggests that experimentally more feasible space-time modulation can model relativistically moving dielectrics \cite{caloz}. Experimentally studying systems in relativistic motion is undoubtedly more challenging.

The question is dual to Doppler metrology, where one employs scattered (or emitted) waves to measure the velocity of a moving object \cite{doppler_metro}. Here we purposefully put the object into motion (or space-time modulate it) to determine its dielectric features. As a result of answering the above question, we have uncovered several effects that may be interesting on their own. 
}

We show that radiating a light on the moving dielectric is not necessary because there is a quantum vacuum response of a photodetector to such a motion. This is a near-field effect (for cylindrical waves), but it can be long-range if the motion is ultra-relativistic; see section \ref{moving}. The vacuum response is also present in the space-time modulated situation, but now the important point is that the response can be detected in the far-field zone (for spherical waves); cf.~section \ref{modulated}. This effect has interesting -- though certainly limited -- analogies with the anomalous Doppler effect; see section \ref{ano}. 

We also studied the scattering of single photons over a space-time modulated dielectric. Here we show explicitly that in the far-field zone of weak scattering the Rayleigh limit for structure determination (along the modulation axis) can be violated, i.e. a better resolution of the internal dielectric structure is possible. 

\comment{
We show that uniform motion or space-time modulation is a resource for dielectric susceptibility. First, there is a vacuum signal that allows us to determine some features of the susceptibility without shining light on the dielectric. Second, when the incident light is there, the classical Rayleigh limit is modified so that the detection is enhanced. 
}

\section*{Acknowledgements}

We were supported by HESC of Armenia, grants 24FP-1F030 and 21AG-1C038.
We thank R. Ghazaryan for discussions on the Doppler effect. 


\bibliography{references}

\appendix

\section{Green's function for the Helmholtz equation}
\label{helmo}

This function is defined as follows:
\begin{align}
\label{sevi}
&
\{(\partial_\sigma\partial_\sigma+\om^2) \delta_{\alpha\beta}-\partial_\alpha\partial_\beta\}G_{\beta\gamma}[\om,\mathbf{r}]
=\delta_{\beta\gamma}\delta(\mathbf{r}),\\
\label{sevo}
& G_{\alpha\beta}[\om>0,\mathbf{r}]=\Big(\delta_{\alpha\beta}+\frac{\partial_\alpha\partial_\beta}{\om^2}\Big) \frac{e^{i\om r}}{(-4\pi) r},
 ~r=|\mathbf{r}|, \\
&\pa_\alpha\equiv\pa/\pa x_\alpha,~~~~\alpha,\beta=1,2,3,\\
&\mathbf{r}=(x_1,x_2,x_3)=(x,y,z), 
\end{align}
where (\ref{sevo}) is found from (\ref{sevi}) using the known expression for the Laplace equation 
\BEA
(\partial_\sigma\partial_\sigma+\om^2) G_\omega(\mathbf{r})=\delta(\mathbf{r}), ~~~~~
G_\omega(\mathbf{r})\equiv \frac{e^{i\om r}}{(-4\pi) r}.
\EEA
\comment{
In text we define tensor of GF%
\begin{align}
G_{\alpha\beta}\left(  x-x^{\prime}\right)  =\left(  \partial_{\alpha}%
\partial_{\beta}-\delta_{\alpha\beta}\partial_{t}^{2}\right)  G\left(
x-x^{\prime}\right) \label{A1}%
\end{align}
where%
\begin{align}
\left(  \Delta-\partial_{t}^{2}\right)  G\left(  x-x^{\prime}\right)
=\delta\left(  x-x^{\prime}\right) \label{A2}%
\end{align}
In text we use explicit expressions of time Fourier transforms of GF (\ref{A1},\ref{A2}):
}

Hence we get from (\ref{sevo})
\begin{align}
\om^2 G_{\alpha\beta}\left[  \omega,\mathbf{r}\right]  &=
\frac{1}{3}\delta_{\alpha\beta}\delta\left(  \mathbf{r}\right)\\
&+
\frac{e^{i\left\vert \omega\right\vert r}}{4\pi r^{3}}\Big\{  \left[
1-i\omega r-\left(  \omega r\right)  ^{2}\right]  \delta_{\alpha\beta}\\
&\qquad\quad-
\left[3-3i\omega r-\left(  \omega r\right)  ^{2}\right]  n_{\alpha}n_{\beta
}\Big\},
\end{align}
$\mathbf{n}=\mathbf{r}/|\mathbf{r}|$.
In the far field limit $\omega r\gg 1$ we get:
\begin{align}
G_{\alpha\beta}\left[  \omega,\mathbf{r}\right]  \rightarrow
G_{\omega}\left(\mathbf{r}\right)\left(  \delta_{\alpha\beta 
}-n_{\alpha}n_{\beta}\right), \label{A6.a}%
\end{align}
while in the near field limit $\omega r\ll 1$ we have
\begin{align}
\omega^2G_{\alpha\beta}\left[  \omega,\mathbf{r}\right]  \rightarrow\frac{1}{3}%
\delta_{\alpha\beta}\delta\left(  \mathbf{r}\right)  +\frac{1}{4\pi r^{3}%
}\left(  \delta_{\alpha\beta}-3n_{\alpha}n_{\beta}\right). \nonumber
\end{align}

\section{Relations with Doppler physics}
\label{doppo}

In the main text, we found the following relation for the scattered field
\begin{align}
E_{\alpha}^{[\text{s}] }[\om,\mathbf{r}] &=-\omega^{2}\int\d\mathbf{r}'\int\d\omega'
G_{\alpha\beta}[\om,\mathbf{r}-\mathbf{r}']\,
\nn&\times
\bar{\varepsilon}[\om-\om',\mathbf{r}'] \,
E_{\beta}^{[\text{in}] }[\om',\mathbf{r}'].
\label{eri2}
\end{align}
Our purpose is to show that this equation directly relates to Doppler physics. Recall that
\begin{align}
\label{hack1}
    \bar\varepsilon[\om,\mathbf{r}]=\int\frac{\d t}{2\pi} e^{it\om}\bar\varepsilon(t,\mathbf{r}).
\end{align}
For $\bar\varepsilon(t,\mathbf{r})$ we select the following simple model of moving non-relativistic dielectric
\begin{align}
\label{hack2}
   \bar\varepsilon(t,\mathbf{r})=\bar\varepsilon(\mathbf{r}-\mathbf{v}t),
\end{align}
where $\mathbf{v}$ is the constant velocity of motion, and 
where for the space-Fourier transform $\hat\varepsilon[\mathbf{k}]$ of the static susceptibility $\bar\varepsilon(\mathbf{r})$ we have
\begin{align}
\label{hack3}
    \bar\varepsilon(\mathbf{r})=\int\d \mathbf{k}\,e^{i\mathbf{k\cdot r}}\, \hat\varepsilon[\mathbf{k}].
\end{align}
Now (\ref{eri2}--\ref{hack3}) imply
\begin{align}
E_{\alpha}^{[\text{s}] }[\om,\mathbf{r}] &=-\omega^{2}\int\d\mathbf{r}'\int\d{\mathbf{k}}
\hat{\varepsilon}[\mathbf{k}] e^{\mathbf{k\cdot r}'} \,\nn
&\times
G_{\alpha\beta}[\om,\mathbf{r}-\mathbf{r}']\, 
E_{\beta}^{[\text{in}] }[\om - \mathbf{k\cdot v} ,\mathbf{r}'].
\label{eri3}
\end{align}
Note that the factor $E_{\beta}^{[\text{in}] }[\om - \mathbf{k\cdot v} ,\mathbf{r}']$ in (\ref{eri3}) has the Doppler-shifted frequency
$\om - \mathbf{k\cdot v}$. 

Let us now assume that 

-- The incident field $E_{\beta}^{[\text{in}] }[\om - \mathbf{k\cdot v} ,\mathbf{r}']$ is classical and concentrated at a frequency $\omega_0>0$. 

-- The motion is along the $x$-axis: $\mathbf{k\cdot v}=k_xv$. 

-- $\hat{\varepsilon}[k_x,k_y,k_z]$ is weakly dependent on $k_y$ and $k_z$ (i.e. the moving dielectric is a thin rod).

These assumptions produce from (\ref{eri3}):
\begin{align}
E_{\alpha}^{[\text{s}] }[\om,\mathbf{r}] \propto \hat{\varepsilon}\Big[\frac{\om-\om_0}{v}\Big]. 
\label{eri4}
\end{align}

\section{Evaluation of vacuum contribution}
\label{evaluation}

\begin{figure}
    \centering    \includegraphics[width=\columnwidth]{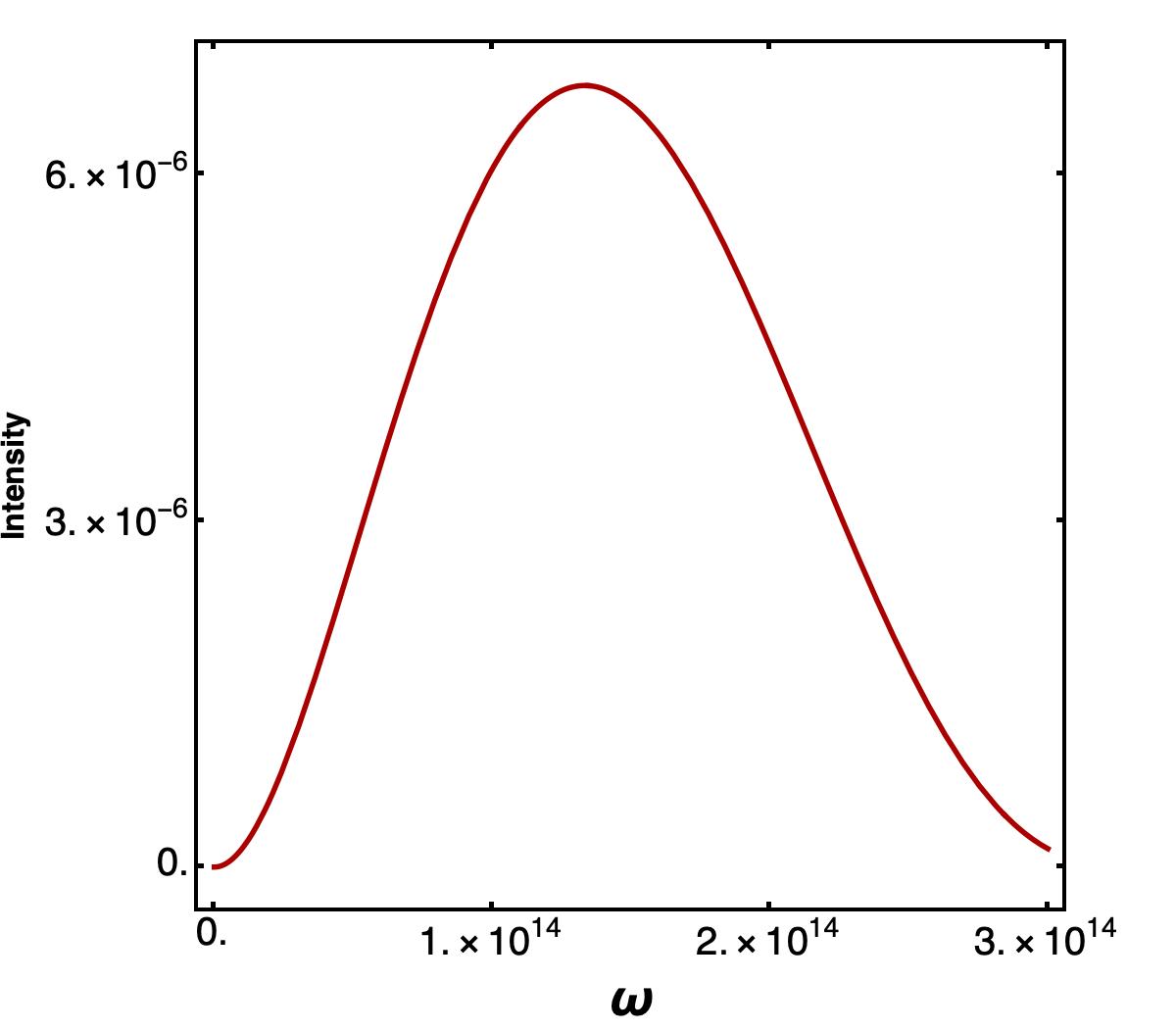}
    \caption{Modulated material in vacuum. The scattered intensity \eqref{vac}, in units of $\frac{\hbar\om^2}{c^3}$, against $\om$~ [Hz]. The parameters are taken as $a=9\times 10^{-6}$cm and $u=0.1 \,c$.}
    \label{figure3}
\end{figure}

We first recover the Planck constant and the speed of light in the temporal Fourier transform of \eqref{Ert}, that is,
\begin{align}
    \mathbf{{E}}^{[\text{in}]  }\left(  \mathbf{r,}t\right)  &=\frac
    {-1}{2\pi}\int \d\mathbf{q}\,\sqrt{\hbar qc} \, \mathbf{e}_{\lambda}\left(  \mathbf{q}\right)
    e^{i\mathbf{q\cdot r}}\\&\times
    \delta(\om-qct){a}_{\lambda}\left(  \mathbf{q}
    \right) +\text{h.c.},\nonumber
\end{align}
and evaluate the scattered intensity \eqref{hope2} of the main text, 
\begin{align}
I[\omega,\mathbf{r};|0\rangle\langle 0|]&=\frac{\om^4\hbar}{u^2c^3(2\pi)^2}\int\d \mathbf{q}\,q\,
\nn&\times
\left|\,\eta\left[-\frac{\om+qc}{u}\right]\,\right|^2 
|\zeta(\mathbf{q,r})|^2,\label{vac_int}
\end{align}
where the function $\zeta$ has the following form in the far field [cf.~(\ref{hope3}, \ref{halagaz})]
\begin{align}
\zeta(\mathbf{q,r})=\frac{e^{i\om \frac{r}{c}}}{r}\int\d\mathbf{r}'
\chi(\mathbf{r}')e^{
\frac{i(\om+qc)x'}{u}-i(\mathbf{q\cdot r}'
+\om \mathbf{n\cdot r'}/c)}.
\label{hag1}
\end{align}
With a model (\ref{hag2}), 
Eq.~(\ref{hag1}) produces
\BEA
&|\zeta(\mathbf{q,r})|^2=
\frac{4 u^2 \chi^2}{r^2}
\frac{\sin^2 \left[\frac{a}{u} (\om+qc-(q_x + \frac{n_x\om}{c}) u)\right]}{\left(\om+qc-(q_x +\frac{n_x\om}{c})u\right)^2}.
\EEA
For the modulation model (\ref{om}) we obtain 
\begin{align}
\label{omni}
&\frac{1}{u^2}    \left|\eta\left[-\frac{\omega+qc}{u}\right]\right|^2\nn
&= \frac{A^2}{4u^2}\left(e^{-A|\frac{\omega+qc}{u}-\frac{1}{a}|}- e^{-A|\frac{\omega+qc}{u}+\frac{1}{a}|}   \right)^2
\end{align}
For a large $A$ (i.e., $A[\frac{\om}{u}+\frac{1}{a}]\gg 1$) we can approximate $\delta(y)\simeq A e^{-2A|y|}$. This leads to (\ref{omni2}). Finally, \eqref{vac_int} evaluates to (\ref{vac}, \ref{f}, \ref{bb}). Fig.~\ref{figure3} presents a numerical illustration of (\ref{vac}) with parameters discussed below.

$f(b,n)$ is slightly smaller when $n_x=0$, i.e., in the direction perpendicular to the modulation direction.

In \eqref{f} the Heaviside theta function selects $b>0$ which means we should have $u>a\om$. For optical frequencies $\om\sim 3\times 10^{14}$ Hz ($\lambda\sim 6\,\mu$m), the condition can be achieved with modulation speeds $u\sim 0.1c$, with $a<10^{-5}$ cm. 
Thus, the dimensionless coefficient in \eqref{vac} is
\begin{align}
   \frac{ \om^2\pi^2 T \chi^2 }{(2\pi)^5 r^2 c a}&=
0.0839884 \label{coefval}
\end{align}
for the parameter values
\begin{align}
u & = 0.1 c;\label{u}\\
a &=  9\cdot 10^{-6} \text{ cm}=0.09\,\mu\text{m};\\
T &= 0.001 \text{ s};\\
\om &=  3\cdot 10^{14} \text{ Hz};\\
\chi &= 10^{-7} \text{ cm}^2;\\
r &= 100 \text{ cm}.
\label{h}
\end{align}
Eq.~(\ref{f}) equals $1.888\cdot 10^{-6}$ for the parameter values given in (\ref{u}-\ref{h}).
Hence, \eqref{vac} is $1.5858\cdot 10^{-7}$ in units of $\frac{\hbar \om^2}{c^3}$; see 
Fig.~\ref{figure3} for illustration.
Note that the linear dimension of the sample $a=0.09\,\mu$m is smaller than the scattered 
field wavelength $\lambda\sim 6\,\mu$m.

\subsection{Comparison of the result with that of single photon and coherent state.}

\subsubsection{Single photon}

To estimate the feasibility of the above estimates, we compare their magnitude with the single photon detection, i.e. we calculate the correlation function of a single free photon. Now recall from (\ref{808})
\begin{align}
\mathbf{{E}}^{[\text{in}]  }\left( \om, \mathbf{r}\right)  &=\frac
{i}{2\pi}\int \d\mathbf{q}\,\sqrt{\hbar qc}\,\mathbf{e}_{\lambda}\left(  \mathbf{q}\right)
e^{i  \mathbf{q\cdot r}}{a}_{\lambda}\left(  \mathbf{q}%
\right)\delta(\om-qc).
\label{111}
\end{align}
We now employ (\ref{111}) with a single photon state of the field [cf.~(\ref{11})]
\begin{align}
\label{abc}
|1\rangle = \int \d^3 k C_\lambda(\bk)a_\lambda(\bk)|0\rangle.
\end{align}
As a simple model for (\ref{abc}), take
\begin{align}
    C_\lambda(\bk) &= \frac{e^{-\frac{k_x^2+k_y^2+k_z^2}{4\sigma^2}}}{\sqrt{2}(2\pi \sigma^2)^{3/4}},\\
    &\sum_\lambda\int \d^3k\left|C_\lambda(\bk)\right|^2=\nn
    & 2\times
    \frac{1}{2(2\pi \sigma^2)^{3/2}}\int \d^3k e^{-\frac{k_x^2+k_y^2+k_z^2}{2\sigma^2}} =1.
\end{align}
Then
\begin{align}
I[\omega,\mathbf{r};|1\rangle\langle 1|]&=
    \left|\int \frac{\d\mathbf{q}}{2\pi}\,\sqrt{\hbar qc}\,\mathbf{e}_{\lambda}\left(  \mathbf{q}\right)
e^{i  \mathbf{q\cdot r}}{C}_{\lambda}\left(  \mathbf{q}\right)\delta(\om-qc)\right|^2
\nn&=
\frac{\hbar \om^2}{c^3}\frac{\om^3}{c^3}\frac{e^{-\frac{\om^2}{2c^2\sigma^2}}} {(2\pi \sigma^2)^{3/2}} \frac{\sin^2(\om r/c)}{(\om r/c)^2}.
\label{1phot}
\end{align}
We choose
\begin{align*}
\sigma &= 0.67~\om/c,
\end{align*}
because this value maximizes (\ref{1phot}). Now \eqref{vac} and \eqref{1phot} are estimated as
\begin{figure}
    \centering
    \includegraphics[width=\columnwidth]{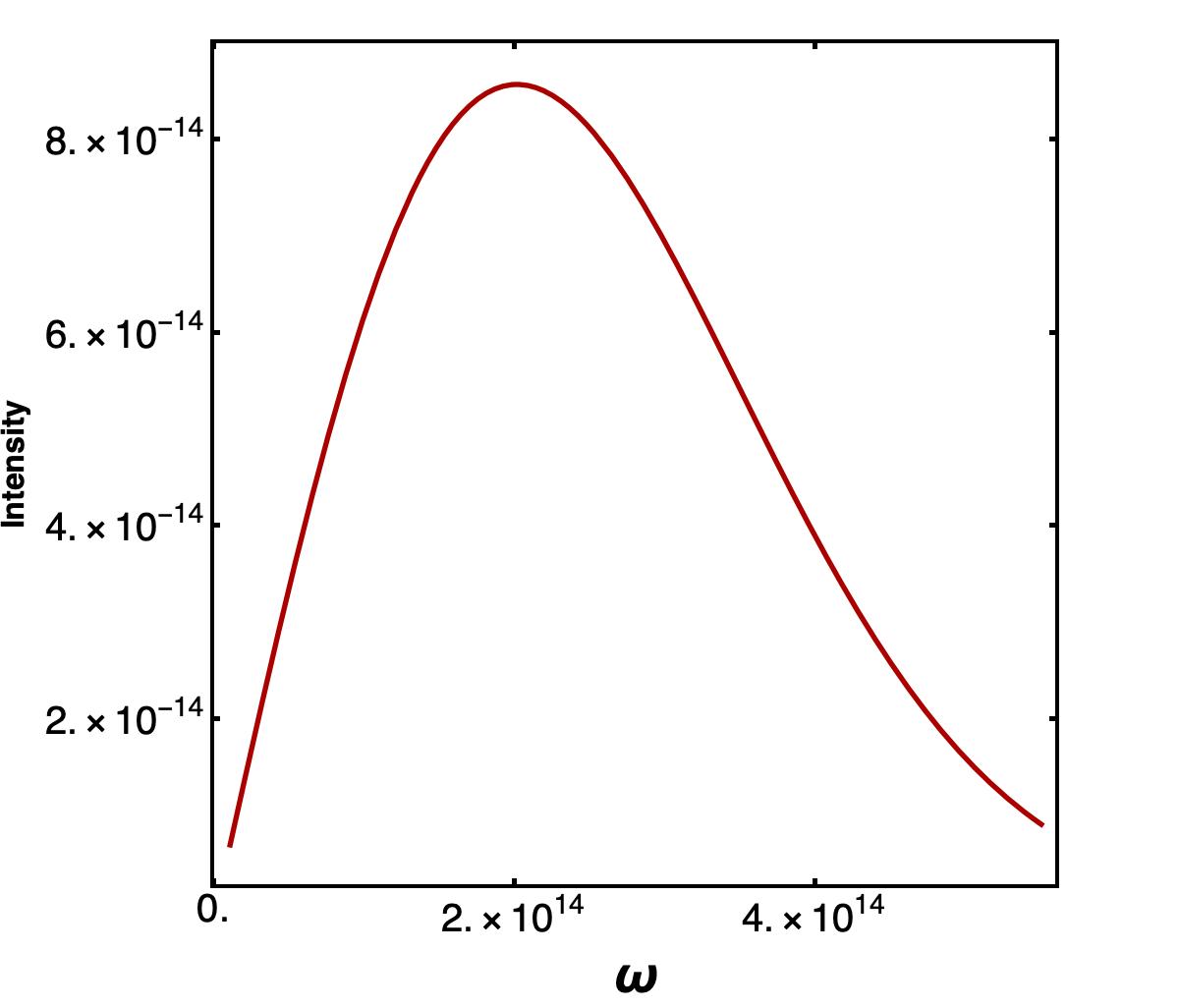}
    \caption{The one-photon intensity \eqref{1phot}, in units of $\frac{\hbar\om^2}{c^3}$, against $\om$~ [Hz], without the $\sin^2$ term. This term was excluded because it just brings fast oscillations of the shown envelope around $1/2$. }
    \label{figure4}
\end{figure}
\begin{align}
\label{darwin}
\eqref{1phot}&\sim  8\times 10^{-15}\,
\times\frac{\hbar \om^2}{c^3} \,\,\text{erg s}^2/\text{cm}^3\\
\eqref{vac}&\sim 1.5\times 10^{-7}\,\times
\frac{\hbar \om^2}{c^3}\,\,\text{erg s}^2/\text{cm}^3.
\label{darwin1}
\end{align}
Hence the photodetection signal from the 
modulated meta-material can be larger than the single-photon signal for a realistic range of parameters. 
The behavior of (\ref{1phot}) is illustrated in Fig.~\ref{figure4}.

\subsubsection{Coherent state of a laser field}

\begin{figure}
    \centering    \includegraphics[width=\columnwidth]{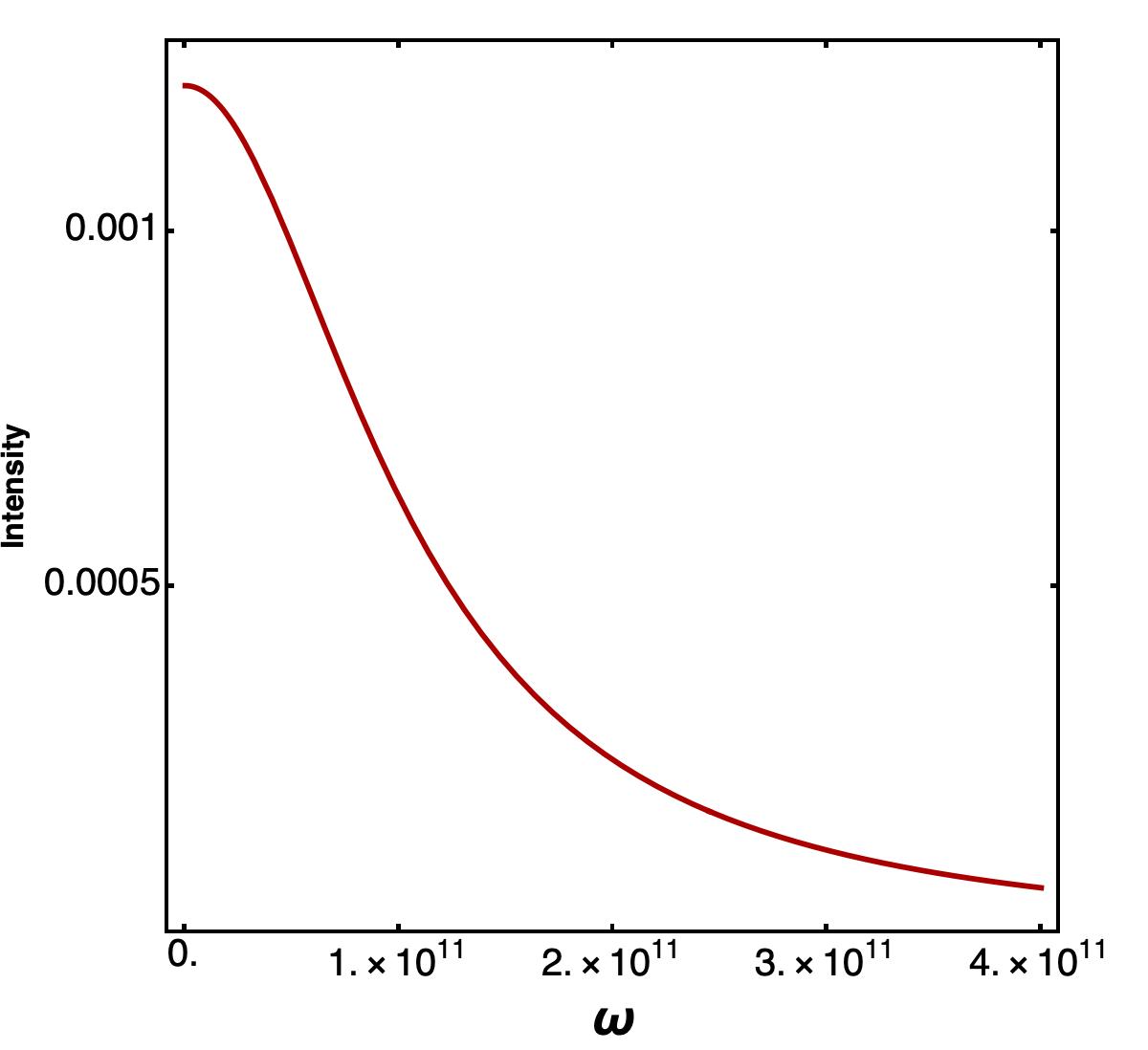}
    \caption{The laser intensity \eqref{laser}, in units of $\frac{\hbar\om^2}{c^3}$, against $\om$~ [Hz].}
\label{laser-fig}
\end{figure}

Let us now compare the estimates (\ref{darwin}, \ref{darwin1}) with the photodetection signal from a coherent state of the field. This corresponds to the laser light. 
Using the continuous-mode representation of the coherent state 
\begin{align}
a_\alpha(\bq)|A_\alpha(\bq)\rangle &= A_\alpha(\bq)|A_\alpha(\bq)\rangle,\\
\langle A_\alpha(\bq)|B_\alpha(\bq)\rangle&=
\exp\Big[-\frac{1}{2}\sum_\alpha\int\d^3q |A_\alpha(\bq)|^2\Big]\nn&\times
\exp\Big[-\frac{1}{2}\sum_\alpha\int\d^3q |B_\alpha(\bq)|^2 \Big]\nn&\times
\exp\Big[\sum_\alpha\int\d^3q A_\alpha^*(\bq) B_\alpha(\bq)\Big]
\end{align}
with
$|A|^2 = \sum_\alpha\int\d^3q|A_\alpha(\bq)|^2$ being the mean photon number. 
Assume that $A_\alpha(\bq)$ is concentrated around $\om\hat{\textbf{z}}/c$, that is, it is a Gaussian with small std $\sigma_z(=\sigma_x=\sigma_y\equiv\sigma)$ and mean $\om/c$ for $q_z$: 
\begin{align}
    A_\alpha(\bq) &= \frac{A}{\sqrt{2}(2\pi \sigma^2)^{3/4}}
    e^{-\frac{q_x^2+q_y^2+(q_z-\om/c)^2}{4\sigma^2}}.
    \label{flambda}
\end{align}
We find for the correlation function
\begin{align}
    I[\omega,\mathbf{r};& |A_\alpha(\bq)\rangle\langle A_\alpha(\bq)|]=
    \frac{\hbar c}{(2\pi)^2}
    \nn&\times
    \Big|
    \int \d^3q\,\sqrt{q}\,  e^{i  \mathbf{q\cdot r}} \delta(\om-qc)A_\alpha(\bq)|A_\alpha(\bq) \rangle\Big|^2 
    \nn&=
    \frac{\hbar|A|^2}{c (2\pi \sigma^2)^{3/2}}  
    e^{-\frac{\om^2}{2c^2\sigma^2}}
    \Big|
    \int_0^\infty \d q\,
    q^{5/2}\,  
    e^{-\frac{q^2}{4\sigma^2}} 
    \nn&\times
    \frac{\sinh(iqr+\frac{\om q}{2c\sigma^2})}{iqr+\frac{\om q}{2c\sigma^2}}
    \delta(q-\om/c)\Big|^2
    \nn&=
    \frac{\hbar \om^2}{c^3} \frac{|A|^2}{ (2\pi)^{3/2}}  
    e^{-\frac{\om^2}{c^2\sigma^2}} \frac{c\sigma/\om}{4 c^2 r^2 \sigma^4/\om^2+1}
    \nn&\times
    \left[\cosh \left(\frac{\om^2}{c^2 \sigma^2}\right)-\cos \left(\frac{2 r \om}{c}\right)\right].\label{laser}
\end{align}  
\comment{\begin{align}
    I[\omega,\mathbf{r};& |A_\alpha(\bq)\rangle\langle A_\alpha(\bq)|]=
    \frac{\hbar c}{(2\pi)^2}
    \nn&\times
    \Big|
    \int \d^3q\,\sqrt{q}\,  e^{i  \mathbf{q\cdot r}} \delta(\om-qc)A_\alpha(\bq)|A_\alpha(\bq) \rangle\Big|^2 
    \nn&=
    \frac{\hbar|A|^2}{c (2\pi \sigma^2)^{3/2}}  
    e^{-\frac{\om^2}{2c^2\sigma^2}}
    \Big|
    \int_0^\infty \d q\,
    q^{5/2}\,  
    e^{-\frac{q^2}{4\sigma^2}} 
    \nn&\times
    \frac{\sinh(iqr+\frac{\om q}{2c\sigma^2})}{iqr+\frac{\om q}{2c\sigma^2}}
    \delta(q-\om/c)\Big|^2
    \nn&=
    \frac{\hbar \om^2}{c^3} \frac{|A|^2}{ (2\pi)^{3/2}}  
    e^{-\frac{\om^2}{c^2\sigma^2}} \frac{c\sigma/\om}{4 c^2 r^2 \sigma^4/\om^2+1}
    \nn&\times
    \left[\cosh \left(\frac{\om^2}{c^2 \sigma^2}\right)-\cos \left(\frac{2 r \om}{c}\right)\right].\label{laser}
\end{align}
}
The behavior of (\ref{laser})
is illustrated on Fig.~\ref{laser-fig}.

We use the previous parameters and $\sigma=0.038\,\om/c$ (recall that $\sigma$ much smaller than $\om/c$), and denoting $\langle n\rangle = |A|^2$ for the mean photon number, we get
\begin{align}
    \eqref{laser} \sim \langle n\rangle \times 10^{-10}\times\frac{\hbar\om^2}{c^3}\,\,\text{erg s}^2/\text{cm}^3.
\end{align}

Summing up, the various contribution to the intensity (spectral density) at the frequency $\om = 3\cdot 10^{14} \text{ Hz}$ in units of $\frac{\hbar \om^2}{c^3}$ are as follows [cf.~(\ref{darwin}, \ref{darwin1})]:
\begin{align*}
    \text{1 photon }&\sim  10^{-15},\\
    \text{coherent } &\sim  \langle n\rangle\times 10^{-10},\\
    \text{vacuum } &\sim 10^{-7}.
\end{align*}
The vacuum scattered intensity is thus equivalent to a laser with a mean photon number of $\langle n \rangle = 10^3$. 
For a laser of 1~mW power (a power for laser pointer), the mean photon number is of order $\langle n\rangle\sim 10^{9}$.

\section{Minkowski's formulation of relativistic continuous medium electrodynamics}
\label{minko}

We start with the standard representation of the electromagnetic field tensor $F^{ik}$ and its dual $F^{*\,ik}$ \cite{landau}:
\BEA
&& F^{ik}=\pa^iA^k-\pa^kA^i,\quad x^i=(t,\boldsymbol{x}) \\
&& F^{*\,ik}=\frac{1}{2}\epsilon^{iklm}F_{lm}, ~~ F^{**\,ik}=-F^{ik},
\EEA
where $A^i$ is the 4-potential, $\pa_i=\pa/\pa x^i$,
and where $\epsilon^{iklm}$ is the Levi-Civita tensor 
with $\epsilon^{0123}=1$. We assume that a continuous medium 
moves along $x^1$-axis with a constant velocity $v$. Hence the 4-velocity $u^i$ reads
in the laboratory frame \cite{landau}:
\BEA
\label{pole}
u^i=\gamma(1,v,0,0), \quad \gamma\equiv (1-v^2)^{-1/2}, \quad u_iu^i=1.
\EEA
We introduce 4-vectors of electric and magnetic field \cite{pal}:
\BEA
\label{4}
E^i=F^{ik}u_k, \qquad B^i=F^{*\,ik}u_k.
\EEA
In the rest frame $u^i=(1,0,0,0)$ we get $E^i=(0,\boldsymbol{E})$, $B^i=(0,\boldsymbol{B})$,
where $\boldsymbol{E}$ and $\boldsymbol{B}$ are the usual electric and magnetic field, respectively. 

Eqs.~(\ref{4}) can be inverted expressing $F^{ik}$ via two anti-symmetric tensors \cite{pal}:
\begin{align}
\label{6x}
&F^{ik}=E^iu^k-E^ku^i-\frac{1}{2}\epsilon^{iklm}(B_lu_m-B_mu_l),\\
\label{61}
&F^{*\,ik}=B^iu^k-B^ku^i+\frac{1}{2}\epsilon^{iklm}(E_lu_m-E_mu_l),\\
\label{62}
&\epsilon^{iklm} B_lu_m=E^iu^k-E^ku^i-F^{ik},
\end{align}
where (\ref{6x}, \ref{61}, \ref{62}) is obtained from (\ref{4}) upon using
standard identities for the Levi-Civita tensor \cite{landau}:
\begin{align}
\epsilon^{iklm}\epsilon_{prlm}&=-2(\delta^i_p\delta^k_r-\delta^i_r\delta^k_p  ),\\
\epsilon^{ikml}\epsilon_{lspq}&=-\delta^i_s(\delta^k_p\delta^m_q-\delta^k_q\delta^m_p)+
\delta^i_p(\delta^k_s\delta^m_q-\delta^k_q\delta^m_s)
\nn&~~~\, -
\delta^i_q(\delta^k_s\delta^m_p-\delta^k_p\delta^m_s).
\end{align}

Consider a moving media that has dielectric response $\ep$ and magnetic
response $\mu$: 
$\ep=\ep(x^i)$ and $\mu=\mu(x^i)$. Once $E^iu^k-E^ku^i$ and
$-\epsilon^{iklm}B_lu_m$ are (resp.) electric and magnetic contributions
to the field tensor [cf.~(\ref{6x})], we define the electromagnetic field
tensor $H^{ik}$ in the medium \cite{hermann,pal}:
\BEA
H^{ik}=\ep(E^iu^k-E^ku^i)-\frac{1}{\mu}\epsilon^{iklm}B_lu_m.
\label{7}
\EEA
$H^{ik}$ combines vectors $\boldsymbol{D}$ and $\boldsymbol{H}$ in the same
way as $F^{ik}$ combines $\boldsymbol{E}$ and $\boldsymbol{B}$.

Note that we can obtain from (\ref{7}) a relation that only contains $\mu$:
\begin{align}
\mu(H_{ik}u_l+H_{li}u_k+H_{kl}u_i)=F_{ik}u_l+F_{li}u_k+F_{kl}u_i.
\end{align}

For $H^{ik}$ Maxwell's equations have a free form \cite{pal}:
\BEA
\label{8}
\pa_iH^{ik}=0.
\EEA
Eqs.~(\ref{8}) can be deduced from a Lagrangian that has a suggestive form \cite{pal}:
\BEA
\label{9}
{\cal L}=\int\d^4 x\, \Big( \ep E_iE^i-\frac{1}{\mu}B_iB^i  \Big),
\EEA
which is to be varied over $A_i$. The equations of motion generated by (\ref{9}) read \cite{pal}:
\BEA
\pa_i \frac{\pa}{\pa [\pa_iA_k]}\Big[ \ep E_iE^i-\frac{1}{\mu}B_iB^i \Big]=0,
\EEA
and coincide with (\ref{7}, \ref{8}).

To work efficiently with (\ref{8}) we invert (\ref{7}) via (\ref{62}):
\BEA
\label{88}
H^{ik}=\frac{1}{\mu}F^{ik}+(\ep-\frac{1}{\mu})(E^iu^k-E^ku^i).
\EEA

We assume a dielectrical situation $\mu=1$ in (\ref{8}, \ref{88}), and find (\ref{11}).

\end{document}